\documentclass[a4paper,fleqn,usenatbib]{mnras}
\usepackage[T1]{fontenc}
\usepackage{ae,aecompl}
\usepackage{graphicx}	% Including figure files
\usepackage{amsmath}	% Advanced maths commands
\usepackage{amssymb}	% Extra maths symbols
\usepackage{txfonts}
\usepackage{multicol}

\newcommand{\reff}[1]{Fig. (\ref{#1})}

\newcommand{\ii}{\mathrm{i}}

\newcommand{\trace}[1]{\mathrm{tr}\left( {#1} \right)}

\newcommand{\de}[1]{\mathrm{d}{#1}}

\newcommand{\Om}{\Omega_{\mathrm{m}}}

\newcommand{\chiH}{\chi_\mathrm{H}}
\newcommand{\chip}{\chi'}

\renewcommand{\vec}[1]{\boldsymbol{#1}}

\newcommand{\dd}{\mathrm{d}}
\newcommand{\bra}{\left\langle}
\newcommand{\ket}{\right\rangle}
\newcommand{\likeli}{\mathcal{L}}
\newcommand{\id}{\mathrm{id}}
\renewcommand{\trace}{\mathrm{tr}}

\graphicspath{{figures/}} % Location of the graphics files

\title[Composite alignment model for IAs]{Angular ellipticity correlations in a composite alignment model for elliptical and spiral galaxies and inference from weak lensing}
\author[T.M. Tugendhat, B.M. Sch{\"a}fer]
{Tim M. Tugendhat\thanks{e-mail: tugendhat@uni-heidelberg.de} and Bj{\"o}rn Malte Sch{\"a}fer\\
Zentrum f\"ur Astronomie der Universit\"at Heidelberg, Astronomisches Rechen-Institut, Philosophenweg 12, 69120 Heidelberg, Germany}

\date{Accepted 2018 January 30. Received 2018 January 07; in original form 2017 September 08}

\pubyear{2017}

\begin{document}
\label{firstpage}
\pagerange{\pageref{firstpage}--\pageref{lastpage}}
\maketitle

% --- abstract --- %
\begin{abstract}
We investigate a physical, composite alignment model for both spiral and elliptical galaxies and its impact on cosmological parameter estimation from weak lensing for a tomographic survey. Ellipticity correlation functions and angular ellipticity spectra for spiral and elliptical galaxies are derived on the basis of tidal interactions with the cosmic large-scale structure and compared to the tomographic weak lensing signal. We find that elliptical galaxies cause a contribution to the weak-lensing dominated ellipticity correlation on intermediate angular scales between $\ell\simeq40$ and $\ell\simeq400$ before that of spiral galaxies dominates on higher multipoles. The predominant term on intermediate scales is the negative cross-correlation between intrinsic alignments and weak gravitational lensing (GI-alignment). We simulate parameter inference from weak gravitational lensing with intrinsic alignments unaccounted; the bias induced by ignoring intrinsic alignments in a survey like Euclid is shown to be several times larger than the statistical error and can lead to faulty conclusions when comparing to other observations. The biases generally point into different directions in parameter space, such that in some cases one can observe a partial cancellation effect. Furthermore, it is shown that the biases increase with the number of tomographic bins used for the parameter estimation process. We quantify this parameter estimation bias in units of the statistical error and compute the loss of Bayesian evidence for a model due to the presence of systematic errors as well as the Kullback-Leibler divergence to quantify the distance between the true model and the wrongly inferred one.
\end{abstract}

% --- keywords --- %

% --- introduction --- % 
\section{Introduction}\label{sect_intro}
Weak gravitational lensing by the cosmic large-scale structure is recognised to be a primary tool for investigating the properties of gravity on large scales through their influence on the expansion dynamics of the Universe and the growth of cosmic structures \citep{mellier_probing_1999, bartelmann_weak_2001, refregier_weak_2003, hoekstra_weak_2008, kilbinger_cosmology_2015}. Weak lensing offers not only sensitivity for the parameters of a dark energy cosmology \citep{huterer_weak_2002, huterer_weak_2010, vanderveld_testing_2012, mortonson_dark_2013}, but also allows the investigation of models of modified gravity \citep{amendola_measuring_2008, dossett_figures_2011, martinelli_constraining_2011} or model independent contraints o the growth- and expansion history \citep{matilla_geometry_2017}, in particular with tomography \citep{hu_power_1999, cooray_power_2001, hu_power_2001, munshi_cosmology_2008, bernstein_comprehensive_2009}. The current generation of tomographic weak lensing surveys has, in conjunction with observations of the cosmic microwave background and galaxy clustering, put strong bounds on dark energy properties, while future surveys set out to constrain the equation of state of dark energy to the percent-level, investigate alternative to general relativity and search for new gravitational phenomena on the largest scales.

A very common assumption in gravitational lensing are intrinsically uncorrelated galaxy shapes and the idea that correlated deflections and deformations of light-bundles caused by gravitational tidal fields of the cosmic-large scale structure are the only mechanisms that introduce correlations in the observed shapes of distant galaxies. This view is challenged by physical interactions of galaxies with their surroundings \citep[see, e.g.][]{jing_intrinsic_2002, forero-romero_cosmic_2014}, most notably gravitational interactions through tidal fields \citep{heavens_intrinsic_2000, croft_weak-lensing_2000, catelan_intrinsic_2001, mackey_theoretical_2002, lee_intrinsic_2005}, giving rise to intrinsic alignments \citep[for reviews, see][]{troxel_intrinsic_2015, joachimi_galaxy_2015, kirk_galaxy_2015, kiessling_galaxy_2015}. 

The alignment of elliptical galaxies is usually described with the linear alignment model, which stipulates that the galaxy's ellipticity depends linearly on the tidal shear field exerted by the large-scale structure on the galaxy \citep{hirata_intrinsic_2004, blazek_testing_2011, blazek_tidal_2015}. The physical picture would be that of a virialised equilibrium system whose potential is gravitationally distorted, leading to a change in shape in the spatial distribution of the dark matter particles and the stars, giving rise to an ellipticity which is proportional to the tidal shear tensor and whose spatial orientation is identical to the principal axis system of the tidal shear \citep{camelio_origin_2015}. In the literature, this alignment model is referred to as the linear alignment model exactly because of this proportionality. Ellipticities of neighbouring galaxies are correlated due to correlations in the tidal fields, which have an identical correlation length as the density field itself.

In spiral galaxies, on the other hand, intrinsic alignments are set at early times prior to gravitational collapse by the generation of angular momentum through tidal torquing. The angular momentum direction is thought to determine the angle of inclination of the galactic disc relative to the line of sight and therefore the measured ellipticity \citep{lee_cosmic_2000, lee_galaxy_2001}. In the accepted tidal torquing picture of angular momentum generation the two relevant quantities are the tidal gravitational shear on one side and the protogalaxy's inertia tensor on the other \citep{bailin_internal_2005}: The product of the two, if their eigensystems do not coincide, will yield a nonzero angular momentum. Angular momentum correlations are quite short-ranged and therefore one would expect ellipticity correlations between spiral galaxies to be only relevant on small scales, and because products of two fields occur in the generation of angular momenta, these alignment models are called quadratic \citep{schaefer_review:_2009, schaefer_angular_2015}. It should be mentioned that there are alternative models for the orientation of disc galaxies that rely on the cosmic vorticity field \citep{libeskind_cosmic_2012} or on anisotropic accretion, and it is common to all models that there can be a rich phenomenology for galaxy alignment \citep{aragon-calvo_spin_2007, pahwa_alignment_2016}.

Apart from intrinsic shape correlations there is an additional effect that the same gravitational tidal shear fields that physically distort an (elliptical) galaxy also give rise to gravitational lensing-induced distortions of a distant background galaxy. As such, this mechanism is able to introduce correlations in the shapes of galaxies that are separated by a very large distance, which is very dissimilar to intrinsic shape correlations which are at most correlated over a physical distance of a few $\mathrm{Mpc}/h$ due to the short range of tidal interaction. As an important consequence, these cross-correlations will be effective across different tomography bins in contrast to the intrinsic alignments themselves, that will only be nonzero within the same tomographic bin.

In the limit of weak lensing the complex gravitational shear is $\gamma$ is added linearly to the complex intrinsic galaxy ellipticity, $\epsilon_\mathrm{obs} = \epsilon + \gamma$ \citep{bartelmann_weak_2001}. Consequently, correlations $\bra\epsilon^{\phantom{\prime}}_\mathrm{obs}\epsilon_\mathrm{obs}^\prime\ket$ between observed ellipticities $\epsilon_\mathrm{obs}$ can be expanded to the intrinsic shape correlation $\bra\epsilon\epsilon^\prime\ket$ (II-correlations), a cross-correlation term $\bra\epsilon\gamma^\prime+\epsilon^\prime\gamma\ket$ containing correlations between the intrinsic ellipticities and the weak lensing shear \citep[GI-terms,][]{hirata_intrinsic_2004}, and finally the genuine lensing signal $\bra\gamma\gamma^\prime\ket$ (GG-correlation) which is dominating for a reasonably deep survey.

Intrinsic alignments have been the target of investigations that quantify their contaminating effect to weak lensing data and the resulting systematic errors in inferring cosmological parameters: Intrinsic alignments of elliptical galaxies are expected to contribute in an intermediate multipole range, whereas the alignments of spiral galaxies resulting from an angular-momentum based model, is only generated on high multipoles, reflecting their intrinsically short-ranged correlation \citep{kirk_impact_2010, kirk_cosmological_2012, schafer_galactic_2012, capranico_intrinsic_2013, merkel_intrinsic_2013, kitching_3d_2014, krause_impact_2016, zieser_cross-correlation_2016}. Similarly, there is an intrinsic alignment influence on cluster number counts \citep{fan_intrinsic_2007}.

Given the large uncertainty about the mechanisms of tidal interaction of galaxies, empirical models have been constructed \citep{schneider_halo_2010} that show a large degree of flexibility and ultimately reduce any systematic errors at the expense of statistical precision. There is as well the possibility of controlling IA-contributions in weak lensing data by methods of self-calibration \citep{zhang_proposal_2010, troxel_self-calibrating_2012,troxel_self-calibration_2012}, by using differences in their redshift dependence \citep{huterer_nulling_2005, joachimi_controlling_2010, joachimi_intrinsic_2010,  shi_controlling_2010},
by different parity of the ellipticity field that is generated by intrinsic alignments and lensing \citep{crittenden_discriminating_2002}, by removing close pairs of galaxies \citep{king_suppressing_2002, heymans_weak_2003, king_cosmic_2005}, by using higher-order statistics \citep{semboloni_effect_2013, merkel_theoretical_2014, munshi_tomography_2014, petri_validity_2016}, or by making use of cross-correlations with other probes of the cosmic large-scale structure \citep{hall_intrinsic_2014, larsen_intrinsic_2016}. To some extent, the survey design can be optimised to have smaller contributions from intrinsic alignments by preferring a deep survey over a shallow one \citep{kirk_optimising_2011}, and it underlines the general trend in cosmology that systematic errors will need to be controlled with much care \citep{amara_optimal_2007, laszlo_disentangling_2012, cardone_power_2014}.

Many of the current generation of weak lensing surveys need to employ a description of the IA-contribution in weak lensing \citep{heymans_weak_2004, joachimi_simultaneous_2010, heymans_cfhtlens_2013, jee_cosmic_2015, de_jong_kilo-degree_2013, 2016arXiv161004606J, des_collaboration_dark_2017}, where alignments seem to be present in elliptical but not in spiral galaxies, and aim to supply suitable parameters from observations at lower redshift \citep{brown_measurement_2002, hirata_galaxy-galaxy_2004, mandelbaum_detection_2006, hirata_intrinsic_2007, joachimi_constraints_2011, mandelbaum_wigglez_2011, lee_intrinsic_2011, joachimi_intrinsic_2013, singh_intrinsic_2015}.

With an improving understanding of galaxy formation and evolution, intrinsic alignments and the relationship between the shape of a galaxy and its environment has been targeted by numerical simulations, who draw a more complicated picture of alignments than the two primary analytical models based on tidal shearing and tidal torquing have in mind. This concerns, e.g. \citet{tenneti_galaxy_2014, tenneti_galaxy_2015, codis_intrinsic_2015, velliscig_intrinsic_2015, chisari_intrinsic_2015, hilbert_intrinsic_2016, chisari_redshift_2016} who point out that many physical processes have an influence on the perceived shape of a galaxy and that alignments can depend on e.g. redshift and luminosity, and that there is in fact an intrinsic shape correlation on the two-point level \citep{tenneti_intrinsic_2015}.

We intend in this paper to combine intrinsic alignments for elliptical and spiral galaxies, and to compute the cross-correlation between the intrinsic shape of elliptical galaxies and the weak gravitational lensing effect, which is, in contrast to spiral galaxies, nonzero for tidal shear fields with Gaussian statistics and has been observed \citep{hirata_intrinsic_2004, okumura_gravitational_2009, hirata_intrinsic_2010}. We chose the formalism for describing intrinsic alignments for both primary galaxy types to be identical in the description of tidal gravitational fields for consistency, and we aim to put the parameter estimation biases that are caused by intrinsic alignments in the interpretation of weak lensing data into a larger context and ask how the degree of belief in a cosmological model is changed by the presence of alignments. In contrast to elliptical galaxies, there is in the case of Gaussian fields no cross-alignment between the intrinsic shapes of elliptical and spiral galaxies, and no GI-type alignment between the intrinsic shape of a spiral galaxy and the weak gravitational lensing effect.

Specifically, the motivation for our investigation was $(i)$ to combine intrinsic alignment models for elliptical and spiral galaxies and to reach, together with the nonzero cross-correlation between the intrinsic shapes of elliptical galaxies and lensing, a complete view of alignments over a wide range of angular scales, $(ii)$ to compute the alignments for elliptical and spiral galaxies from the same fundamental formalism and to reach a high degree of consistency in our description, $(iii)$ to estimate their parameter biasing effect on a topical dark energy cosmology in inference from tomographic weak lensing spectra, and $(iv)$ to quantify this biasing effect not only in terms of systematical errors, but also to compute the loss in Bayesian evidence and the Kullback-Leibler distance between the true and the wrongly inferred likelihood: This will be of particular importance as future surveys will be limited in their parameter-constraining and model selecting power by systematical rather than statistical errors. We carry out our investigation of intrinsic alignment and its interplay with weak gravitational lensing for the Euclid\footnote{\url{http://sci.esa.int/euclid/}} weak lensing survey and compute alignment correlation functions and spectra and their cross-correlation with lensing for a tomographic survey with Euclid's characteristics, and intend to address questions in relation to the information content of weak lensing surveys \citep{carron_probe_2011}.

After the introduction in Sect.~\ref{sect_intro} we recapitulate key concepts of cosmology and gravitational lensing in Sect.~\ref{sect_cosmology}, followed by the theory of the intrinsic alignments in Sect.~\ref{sect_alignments}. We present our results on ellipticity correlations for our composite alignment model in Sect.~\ref{sect_ellipticity} before discussing statistical errors in Sect.~\ref{sect_statistical} and systematical errors in Sect.~\ref{sect_systematical}, with particular emphasis on the Euclid-mission. We summarise and discuss our findings in Sect.~\ref{sect_conclusion}. The reference cosmology for our investigation is a $w$CDM-cosmology with a non-clustering dark energy component with an equation of state $w$. Specific choices for the cosmological parameters are $\Omega_\mathrm{m} = 0.32$, $\sigma_8 = 0.83$, $h = 0.68$, $n_s = 0.96$ and $w = -1$, which are motivated by Planck's results \citep{2015arXiv150201590P}.

We assume Gaussian statistics for fluctuations in the distribution of matter, and assume that this is approximately valid even in the regime of nonlinear structure formation. Angular-momentum based alignments will be computed from a linear CDM-spectrum because angular momentum generation is a perturbative process in the early stages of galaxy formation, whereas tidal shears distorting an elliptical galaxy will be derived from a nonlinear CDM-spectrum in the same way as the gravitational lensing effect. Consequences of the assumption of Gaussianity in the statistics of tidal shear fields will be vanishing GI-correlations for spirals and vanishing cross-correlations between the shapes of spiral and elliptical galaxies.

% --- cosmology --- %
\section{Gravitational lensing and cosmology}\label{sect_cosmology}

% ---  --- %
\subsection{Distances and structure growth}
The comoving distance $\chi$ corresponding to the scale factor $a$ is obtained by
\begin{equation}
\chi = c\int_a^1\frac{\dd a}{a^2H(a)}
\end{equation}
with the Hubble-function $H(a)$,
\begin{equation}
\frac{H^2(a)}{H_0^2} = \frac{\Omega_\mathrm{m}}{a^3} + \frac{1-\Omega_\mathrm{m}}{a^{3(1+w)}},
\end{equation}
for a spatially flat FLRW-cosmology with a constant dark energy equation of state parameter $w$, with the Hubble-distance $\chi_H=c/H_0$ as the natural distance scale. The linear structure growth in these cosmologies is determined by the growth equation,
\begin{equation}
\frac{\dd^2 D_+}{\dd a^2} + \frac{1}{a}\left(3+\frac{\dd\ln H}{\dd\ln a}\right)\frac{\dd D_+}{\dd a} - \frac{3}{2a^2}\Omega_\mathrm{m}(a)D_+(a) = 0
\end{equation}
which yields the growth function $D_+(a)$ as the growing solution, such that the density contrast acquires a time evolution according to $\delta(\bmath{x},a) = D_+(a)\delta(\bmath{x})$, for $\left|\delta(\bmath{x},a)\right|\ll 1$. The statistics of the density contrast is determined by the CDM-spectrum, $\bra\delta(\bmath{k})\delta(\bmath{k}^\prime)\ket = (2\pi)^3\delta_D(\bmath{k}+\bmath{k}^\prime) P_\delta(k)$, which is normalised the variance $\sigma_8^2$ on the scale of $R = 8~\mathrm{Mpc}/h$,
\begin{equation}
\sigma_8^2 = \int\frac{k^2\dd k}{2\pi^2}\:\left(\frac{3}{kR}j_1(kR)\right)^2\:P_\delta(k),
\end{equation}
where we use a transfer function from \cite{1986ApJ...304...15B} for the linear CDM-spectrum $P_\delta(k)$ and \cite{Smith:2002dz} for the nonlinear CDM-spectrum. We would like to point out that the nonlinear spectrum will be used for the weak gravitational lensing effect \citep{casarini_non-linear_2011}, for the alignments of elliptical galaxies and the cross-correlation between the shape of ellipticals and weak lensing, because these three effects probe tidal gravitational fields in the evolved large-scale structure. Alignments of spiral galaxies are set through tidal torquing as a perturbative process from the initial conditions of structure formation, and therefore we use a linear CDM-spectrum to compute their alignment.

From the spectrum of the density contrast $\delta$ one can obtain that of the gravitational potential $\Phi$ in units of $c^2$ by substituting the comoving Poisson-equation, $\Delta\Phi = 3\Omega_\mathrm{m}/(2\chi_H^2)\delta/a$,
\begin{equation}
P_\Phi(k,a) = \left(\frac{3\Omega_\mathrm{m}}{2a}\right)^2 \frac{P_\delta(k,a)}{(\chi_Hk)^4},
\end{equation}
with the Hubble-distance $\chi_H$ making the relation dimensionless and with a scaling $P_\Phi(k,a)\propto (D_+/a)^2$ (in the limit of linear structure formation), which is relevant for gravitational lensing and intrinsic alignments alike, as both are gravitational interactions. For the computation of tidal shear fields in the context of intrinsic alignments, we impose a Gaussian smoothing,
\begin{equation}
P_\Phi(k) \rightarrow P_\Phi(k) \exp\left(-(kR)^{2}\right)
\label{eq:Psmooth}
\end{equation}
on the mass scale $M$ of the galaxy, i.e. on a spatial scale determined by $M = 4\pi/3\:R^3\Omega_\mathrm{m}\rho_\mathrm{crit}$ \citep{crittenden_spin-induced_2001, schaefer_angular_2015}. We choose a typical value $M=10^{12}M_\odot/h$ for this mass scale, which corresponds to the dark matter mass of a galaxy similar to the Milky Way.

% ---  --- %
\subsection{Gravitational Lensing}
Weak gravitational lensing refers to the change in the cross-sectional shape of light bundles caused by gravitational tidal fields of the cosmic large-scale structure: Ellipticities $\epsilon$ of distant galaxies get mapped in the regime of weak lensing, where shear $\gamma$ and convergence $\kappa$ are small, $|\gamma|, |\kappa|\ll 1$, according to the transformation $\epsilon\rightarrow\epsilon+\gamma$.

The lensing potential $\psi_i$ is given by a projection integral, where we introduce the index $i$ to refer to the lensing potential from the $i$th tomographic bin,
\begin{equation}
\psi_i = \int_0^{\chi_H}\dd\chi\:W_i(\chi)\Phi,
\label{eqn_lensing_potential}
\end{equation}
relating $\psi_i$ to the gravitational potential $\Phi$ through weighting function $W_i(\chi)$,
\begin{equation}
W_i(\chi) = 2\frac{D_+(a)}{a}\frac{G_i(\chi)}{\chi}.
\label{eq:lensingweight}
\end{equation}
As a line of sight-integrated quantity, the projected potential contains less information than the sourcing field $\Phi$. In tomography, one defines the tomographic lensing efficiency function $G_i(\chi)$,
\begin{equation}
G_i(\chi) = \int^{\chi_{i+1}}_{\mathrm{min}(\chi,\chi_i)}\dd\chip\: 
n(\chip) \frac{\dd z}{\dd\chip}\left(1-\frac{\chi}{\chip}\right),
\end{equation}
with $\dd z/\dd\chip = H(\chip) / c$ and the bin edges $\chi_i$ and $\chi_{i+1}$, respectively. A common choice of the redshift distribution $n(z)\dd z$ used in forecasts for Euclid is the parameterisation,
\begin{equation}
n(z)\dd z \propto \left(\frac{z}{z_0}\right)^2\exp\left[-\left(\frac{z}{z_0}\right)^\beta\right]\dd z,
\end{equation}
with $\beta=3/2$ causing a slightly faster than exponential decrease at large redshifts \citep{laureijs_euclid_2011}.

Combining all results one obtains the angular spectra $C_{\psi,ij}(\ell)$ of the tomographic weak lensing potential $\psi_i$ in the flat-sky approximation \citep{limber_analysis_1954},
\begin{equation}
C^\psi_{ij}(\ell) = \int_0^{\chi_H}\frac{\dd\chi}{\chi^2}\:W_i(\chi)W_j(\chi)\:P_\Phi(k=\ell/\chi).
\end{equation}
Weak lensing convergence is related to the lensing potential through the relationship $\Delta\psi = 2\kappa$, therefore its spectrum $C^\kappa_{ij}(\ell)$ is equal to $\ell^4C^\psi_{ij}(\ell)/4$, which is identical to the $E$-mode spectrum $C^{\gamma}_{E,ij}(\ell)$ of the weak lensing shear.

Observed spectra of the weak lensing shear will contain a constant shape noise contribution $\sigma_\epsilon^2 n_\mathrm{bin}/\bar{n}$,
\begin{equation}
\hat{C}^{\gamma}_{E,ij}(\ell) = 
C^{\gamma}_{E,ij}(\ell) + \sigma_\epsilon^2\frac{n_\mathrm{bin}}{\bar{n}}\delta_{ij},
\end{equation}
under the assumption of uncorrelated shape noise. The spectra $C^{\gamma}_{E,ij}(\ell)$ are nonzero for $i\neq j$ leading to a non-diagonal covariance matrix: This will be notably different in the case of intrinsic alignments, which, as local effects, are only correlated within the same tomography bin. Cross-correlations between intrinsic alignments and weak lensing, however, will again be correlated across tomography bins due to lensing's being an integrated effect. We chose the bin edges in a way that they contain identical fractions of the total number of galaxies $4\pi f_\mathrm{sky}\bar{n}$, with the number density per solid angle $\bar{n}$ and the sky coverage $f_\mathrm{sky}$, as the details of the tomography bin choice are do not strongly affect forecasts on statistics.

% --- intrinsic alignments --- %
\section{Intrinsic Alignments}\label{sect_alignments}
Correlations in ellipticities of galaxies are to a large part generated by the weak gravitational lensing effect, but galaxies can have, due to a number of mechanisms relation to their formation process and their gravitational interaction, correlated intrinsic shapes: In the case of spiral galaxies, one could explain shape correlations in terms of correlated orientations of galactic discs, which reflect similarities in their angular momentum generation by tidal torquing. Elliptical galaxies, on the other hand, are thought to reflect in their shape the magnitude and orientation of the tidal gravitational field that is exerted by the cosmic large-scale structure. Gravitational lensing and intrinsic alignments can be expected to show different physical and statistical characteristics, but whether a complete separation is feasible is yet unknown. Instead, we aim in this investigation to construct a composite alignment model as a contaminant of the weak lensing signal, and propagate this contamination to a forecast of systematical errors that could be expected Euclid's weak lensing survey, if intrinsic alignments are not modelled, mitigated or removed. We build our alignment models on the established mechanisms of tidal shearing for elliptical galaxies and tidal torquing for spiral galaxies, and combine their spectra with a fixed fraction of $q=0.7$ of spiral galaxies. This choice is only a rough approximation of the intricate relationships between galaxy morphologies, age, and local galaxy density. The implication of an evolving $q(z)$ would go beyond the scope of this paper, which is focusing on model simplicity and is mainly concerned with the impact of the alignment mixture on a lensing survey. A factor of $q$ is needed to calculate the relative amplitudes that both models contribute to the total alignment signal. Since, however, the change of the spiral fraction with respect to redshift seems to be weak for low-to-intermediate redshifts \citep[see e.g.][]{dressler_morphdens_1997}, a constant $q$ is a valid assumption. Furthermore, replacing $q$ with $q(z, \delta, \dots)$ is easily achieved in this model and can be subject of future investigation.\\
Both intrinsic alignment models are in their core models of tidal gravitational interaction of galaxies with the ambient large-scale structure.

% ---  --- %
\subsection{Correlations in the tidal shear field}\label{tidal_shear_correlations}
Our alignment models are both based on tidal interaction of galaxies with the cosmic large-scale structure. As such, they require the correlation function of tidal shear tensors \citep{catelan_correlations_2001}, on which the correlation of ellipticities will be modelled. Correlations
\begin{equation}
C_{\alpha\beta\gamma\delta}(r)\equiv\bra\Phi_{\alpha\beta}(\bmath{x})\Phi_{\gamma\delta}(\bmath{x}^\prime)\ket
\end{equation} 
of the tidal shear components
\begin{equation}
\Phi_{\alpha\beta}(\bmath{x}) = \frac{\partial^2}{\partial x_\alpha\partial x_\beta}\Phi(\bmath{x})
\end{equation}
as a function of distance $r=\left|\bmath{x}-\bmath{x}^\prime\right|$, can be computed to be
\begin{equation}
\begin{split}
C_{\alpha\beta\gamma\delta}(r) = &
(\delta_{\alpha\beta}\delta_{\gamma\delta}+\delta_{\alpha\gamma}\delta_{\beta\delta}+\delta_{\alpha\delta}\delta_{\beta,\gamma})\zeta_2(r)+\\ 
&(\hat{r}_\alpha \hat{r}_\beta \delta_{\gamma\delta}+\mathrm{5~perm.}) \zeta_3(r)+\\
&\hat{r}_\alpha \hat{r}_\beta \hat{r}_\gamma \hat{r}_\delta \zeta_4(r),
\end{split}
\label{eqn_decomp}
\end{equation}
and to involve the $\zeta_n(r)$-functions, 
\label{sec:quadalign}
\begin{equation}
\zeta_n(r) = \left(-1\right)^n r^{n-4}\int\frac{\de{k}}{2\pi^2}\:P_\Phi(k)\,k^{n+2}\,j_n(kr),
\label{eq:zeta_n}
\end{equation}
as Fourier-transforms the spectrum $P_\Phi(k)$ and its derivatives under the assumption of isotropy \citep{crittenden_spin-induced_2001}. $\hat{r}_\alpha$ is the $\alpha$-component of the unit vector parallel to $r=\bmath{x}-\bmath{x}^\prime$.

In the context of tidal interaction of galaxies we will focus on the change in shape due to tidal shears in the case of ellipticals and the orientation of the angular momentum given by tidal torquing. As both modes of tidal interaction do not depend on the trace of the tidal shear, it is reasonable to define the correlation function $\tilde{C}_{\alpha\beta\gamma\delta}(r)$ of the traceless tidal shear $\tilde{\Phi}_{\alpha\beta} = \Phi_{\alpha\beta} - \Delta\Phi/3\:\times\delta_{\alpha\beta}$,
\begin{equation}
\begin{split}
\tilde{C}_{\alpha\beta\gamma\delta}(r) = &
C_{\alpha\beta\gamma\delta}(r) \\
&- \frac{1}{3}\left(\delta_{\gamma\delta}\left(5\zeta_2(r)+\zeta_3(r)\right) + \hat{r}_\gamma \hat{r}_\delta\left(7\zeta_3(r)+\zeta_4(r)\right)\right)\delta_{\alpha\beta} \\
&- \frac{1}{3}\left(\delta_{\alpha\beta}\left(5\zeta_2(r)+\zeta_3(r)\right) + \hat{r}_\alpha \hat{r}_\beta\left(7\zeta_3(r)+\zeta_4(r)\right)\right)\delta_{\gamma\delta} \\
&+ \frac{1}{9}\left(15\zeta_2(r)+10\zeta_3(r)+\zeta_4(r)\right)\delta_{\alpha\beta}\delta_{\gamma\delta}.
\end{split}
\end{equation}
The trace of the tidal shear matrix, on the other hand, which is proportional to the local density through Poisson's equation, would cause galaxies to be more compact, which has led to the idea of intrinsic magnification \citep{heavens_combining_2013, huff_magnificent_2014}. Although this effect is weak and difficult to measure, it would be very interesting from a fundamental point of view because it would constitute a test of the locality of the gravitational field equation because it would relate the isotropic part of the tidal shear tensor, which causes a change in size of the galaxy, to the local value of the density $\delta$ \citep{heavens_cosmic_2011, alsing_weak_2014, ciarlariello_modelling_2016}.

% ---  --- %
\subsection{Alignments of elliptical galaxies: tidal shearing-model}
The model for elliptical galaxies is that of a virialised system in which stars perform random motion with a constant velocity dispersion $\sigma^2$ and anisotropy parameter $\beta$ in a gravitational potential $\Phi$. The density $\rho$ of stars follows the Jeans-equation \citep[see][]{2008gady.book.....B},
\begin{equation}
\frac{1}{\rho}\frac{\partial}{\partial r}(\rho\sigma^2) + \frac{2\beta}{r}\sigma^2 = 
-\frac{\partial\Phi}{\partial r},
\end{equation}
which results in a dependence $\rho\propto\exp(-\Phi/\sigma^2)$ for isotropic systems, $\beta = 0$. A weak distortion of the gravitational field due to tidal shears exerted by the large-scale structure can be implemented by adding a quadrupole field to the potential $\Phi(\bmath{x})$ centered on the potential minimum at centre of mass of the galaxy $\bmath{x}_0 = 0$,
\begin{equation}
\Phi(\bmath{x}) \rightarrow 
\Phi(\bmath{x}) + \frac{1}{2}\frac{\partial^2\Phi(\bmath{x}_0)}{\partial x_\alpha\partial x_\beta}x_\alpha x_\beta.
\end{equation}
Consequently, the density of particles would change according to
\begin{equation}
\rho \propto 
\exp\left(-\frac{\Phi(\bmath{x})}{\sigma^2}\right)
\times
\left(1-\frac{1}{2\sigma^2}\frac{\partial^2\Phi(\bmath{x}_0)}{\partial x_\alpha\partial x_\beta}x_\alpha x_\beta\right),
\label{eqn_jeans}
\end{equation}
which in turn gives rise to an ellipticity that is linear in the tidal shear field $\partial^2\Phi$, and we will refer to this constant of proportionality as $D$. Eq.~\ref{eqn_jeans} suggests that the ellipticity measures tidal shear in units of $\sigma^2/R^2$ with the size of the galaxy $R$, and it can be argued that $\sigma^2/R^2$ constant \citep{piras_mass_2017}: In virial equilibrium, one would expect the kinetic specific energy in unordered motion $\sigma^2$ to be equal to the specific potential energy $GM/R$ and the mass $M$ to be proportional to $R^3$, such that $\sigma^2/R^2 = GM/R^3 = \mathrm{const}$, and for that reason we would not expect $D$ to be dependent on mass or redshift: In fact, the ellipticity is purely determined by the magnitude and orientation of the external tidal shear fields. We would like to point out, however, that this picture assumes that elliptical galaxies are effectively test particles that react to the external tidal shear field and do not otherwise interact with their environment, which would be notably different in clusters of galaxies \citep{hao_intrinsic_2011}. It is remarkable that intrinsic alignments formally depend on the ratio between the tidal shear and the velocity dispersion $\sigma^2$, whereas gravitational lensing measures tidal shear in units of $c^2$ as the natural velocity scale.

Collecting these results, the linear alignment model for elliptical galaxies sets the complex ellipticity $\epsilon = \epsilon_+ + \epsilon_\times$ into relation with the components of the external tidal shear,
\begin{equation}
\epsilon = D \left(\frac{\partial^2\Phi}{\partial x^2} - \frac{\partial^2\Phi}{\partial y^2} + 2\ii \frac{\partial^2\Phi}{\partial x\partial y}\right),
\label{eq:linearmodel}
\end{equation}
if one assumes a coordinate system in which $x$ and $y$ are oriented perpendicular to the line of sight. In this way, ellipticity correlations between galaxies separated by $\vec{r}$ are traced back to traceless tidal shear correlations, expressed in a fixed coordinate system where the $z$-axis is oriented along the line of sight, i.e. where the vector $\bmath{r}$ has the components $(r\sin\alpha,0,r\cos\alpha)$, and one obtains using the results of Sect.~\ref{tidal_shear_correlations}:
\begin{align}
\left\langle\epsilon_+ \, \epsilon_+'\right\rangle (\vec{r})&= D^2\,\left(4\,\zeta_2(r)+4\,\sin^2(\alpha)\,\zeta_3(r)+\sin^4(\alpha)\,\zeta_4(r)\right),\\
\label{eq:C3Dpp}
\left\langle\epsilon_{\times} \, \epsilon_{\times}'\right\rangle(\vec{r}) &= 4\,D^2\,\left(\zeta_2(r)+\sin^2(\alpha)\,\zeta_3(r)\right).
\end{align}
For completeness, we relate the autocorrelation of the absolute value $\epsilon_s=\sqrt{\epsilon_+^2+\epsilon_\times^2}$ and its cross-correlation to the $\epsilon_+$-component to the tidal shear field,
\begin{align}
\left\langle\epsilon_s \, \epsilon_s'\right\rangle (\vec{r})&= D^2\,\left( 8 \, \zeta_2(r) + 8 \sin^2(\alpha) \, \zeta_3(r) + \sin^4(\alpha)\,\zeta_4(r) \right),\\
\label{eq:C3Dss}
\left\langle\epsilon_s \, \epsilon_+'\right\rangle(\vec{r}) &= -D^2\,\left(6 \, \sin^2(\alpha)\,\zeta_3(r) + \sin^4(\alpha)\,\zeta_4(r)\right),
\end{align}
while the two remaining combinations $\left\langle\epsilon_s\,\epsilon_\times^\prime\right\rangle$ and $\left\langle\epsilon_+\,\epsilon_\times^\prime\right\rangle$ can be shown to be zero.

Due to the fact that the observed ellipticities are linear in the tidal shear components, one expects correlation functions to scale $\propto D^2$ and to reflect tidal shear correlations through the $\zeta_n(r)$-functions. We have chosen a description in real space to be in our formalism as similar as possible to the description of alignments of spiral galaxies, which we will discuss in Sect.~\ref{sect_torquing}, but point out that Fourier-approaches which link modes of the ellipticity field directly to the CDM-spectrum $P_\delta(k)$ in a fashion similar to lensing are likewise possible and effectively equivalent. Second-order corrections due to clustering and peculiar velocities can likewise be included by convolution of the respective fields \citep{blazek_tidal_2015}. 

We assume that elliptical galaxies react instantaneous to external tidal shear fields, and consequently, have the ellipticity scale in proportionality to $D_+/a$. We would argue that this is a good approximation because galaxies would react to changes in the gravitational potential on a time scale equivalent to the free-fall time which is of the order $1/\sqrt{G\rho} \simeq 0.373/H_0$ for a value of $\rho=200\Omega_\mathrm{m}\rho_\mathrm{crit}$ with $\rho_\mathrm{crit} = 3H_0^2/(8\pi G)$. This time scale is short compared to the Hubble-time scale $1/H_0$ on which cosmic structure formation takes place and the growth function $D_+$ changes: Due to the gravitational interaction, the relevant quantity is $D_+/a$ which is almost constant and only decreases slightly in dark energy cosmologies at late times, at least in linear structure formation. Whether in fact the ellipticity can be used to provide a measurement of the tidal field strength to a time earlier than the observation and hence outside the past light cone, would depend on the detailed understanding of the alignment parameter $D$ and the dynamics of the gravitational tidal interaction. Despite these arguments we will assume that elliptical galaxies react instantaneously to external tidal shear fields, and maintain this assumption even if these fields are generated by nonlinearly evolving structures, even though their dynamical time scale might then be even shorter. Different cases of tidal interaction of virialised structures are discussed in \citep{camelio_origin_2015}.

In comparison other studies of alignment models for elliptical galaxies with tidal shearing as the interaction mechanism \citep{blazek_separating_2012, blazek_testing_2011, blazek_tidal_2015} our model is deliberately chosen to be set up in real space instead of configuration space in order to use the same formalism and to be consistent to the alignment model for spiral galaxies. In our line of sight-projection we do not incorporate clustering effects or redshift-space distortions due to peculiar motion, but we would argue that in particular the latter should not matter due to the wide redshift bins of Euclid's weak lensing. But clustering of the sources would generate a non-Gaussian statistical signal, which we ignore as it would interfere with our statistical analysis in Sect.~\ref{sect_systematical}, which assumes Gaussian statistics, and as it would source cross-correlations between the shapes of different galaxy types. It should be possible, however, to compute all shape correlations, including those of spiral galaxies, consistently in Fourier-space as well, if our model is replaced by that of \citet{mackey_theoretical_2002} or ultimately \citep{blazek_beyond_2017}. 

Applying a Limber projection in real space \citep{limber_analysis_1954, loverde_extended_2008}, we find for the angular correlation functions:
\begin{equation}
\left\langle\epsilon_{a,i} \, \epsilon_{a,i}'\right\rangle(\theta) = 
\int_{0}^{\chiH}\de\chi\: n_i(\chi)\:
\int_{\chi_{i}}^{\chi_{i+1}}\de\chi'\: n_{i}(\chi') 
\left\langle\epsilon_{a} \, \epsilon_{a}'\right\rangle\Big(\vec{r}(\chi, \chi')\Big)
\end{equation}
where $a \in \left\{+,\times\right\}$ and $i$ denotes the $i$th tomographic bin.

Note that the lower limit of the $\chi'$-integration is not limited by the outer integration parameter $\chi$. This expresses the fact that the aligned structures do not necessarily need to be behind the affecting fields from the observer's point of view, which is a clear distinction from weak lensing (eq.~\ref{eq:lensingweight}). Defining 
\begin{equation}
C^{\epsilon,\text{II}}_{\pm,i}(\theta) = \left\langle\epsilon_{+,i} \, \epsilon_{+,i}'\right\rangle \pm \left\langle\epsilon_{\times,i} \, \epsilon_{\times,i}'\right\rangle,
\label{eq:Cpmi}
\end{equation}
we can calculate the ellipticity $E$- and $B$-mode spectra of linear alignment $C_{i}^{\epsilon}(\ell)$ via a Fourier transform,
\begin{align}
C^{\epsilon,\text{II}}_{{E},i}(\ell)&=\pi \int \theta \de \theta ~\Big(C^{\epsilon,\text{II}}_{+,i} (\theta) \, J_0(\ell\theta)+C^{\epsilon,\text{II}}_{-,i}(\theta)J_4(\ell\theta) \Big),\\
C^{\epsilon,\text{II}}_{{B},i}(\ell)&=\pi \int \theta\de \theta ~\Big(C^{\epsilon,\text{II}}_{+,i} (\theta) \, J_0(\ell\theta)-C^{\epsilon,\text{II}}_{-,i}(\theta)J_4(\ell\theta) \Big),
\end{align}
by Fourier-transform and isolate the $E$- and $B$- parity eigenmodes of the intrinsic ellipticity fields.

% ---  --- %
\subsection{Cross-correlation between alignments and weak lensing}
Unlike in the quadratic model, we can find nonzero GI-contributions. GI-correlations are cross-correlations between weak lensing and intrinsic alignments, which in physical terms means that a structure affects the alignment of nearby galaxies whilst lensing the light from background galaxies.

Since the ellipticities $\gamma$ and $\epsilon$ only differ by a sign for the same tidal shear field (eq.~ \ref{eq:linearmodel}), the 3d correlation functions are very similar to the II-case, namely
\begin{equation}
\begin{aligned}
&\left\langle\gamma_+ \, \epsilon_+'\right\rangle (\vec{r}) = \\
&-\int_{0}^{\chiH}\hspace{-2mm}\de \chi ~ W(\chi)\,D\left(4\,\zeta_2(r)+4\,\sin^2(\alpha)\,\zeta_3(r)+\sin^4(\alpha)\,\zeta_4(r)\right) = \\
&-\int_{0}^{\chiH}\hspace{-2mm} \de \chi ~ \frac{W(\chi)}{D}\,\left\langle\epsilon_+ \, \epsilon_+'\right\rangle (\vec{r}),\\
\end{aligned}
\end{equation}
and correspondingly,
\begin{equation}
\begin{aligned}
&\left\langle\gamma_\times \, \epsilon_\times'\right\rangle (\vec{r}) = \\
&4 \int_{0}^{\chiH} \de \chi ~W(\chi)\,D\,\left(\zeta_2(r)+\sin^2(\alpha)\,\zeta_3(r)\right) = \\
&\int_{0}^{\chiH} \de \chi \frac{W(\chi)}{D}\,\left\langle\epsilon_\times \, \epsilon_\times'\right\rangle (\vec{r}).\\
\end{aligned}
\end{equation}
The 2d correlation functions are then, analogous to the II-case, 
\begin{equation}
\left\langle\gamma_{a,i} \, \epsilon_{a,j}'\right\rangle(\theta) =
\mp\int_{0}^{\chiH} \hspace{-2mm}\de \chi \int_{\chi_{j}}^{\chi_{j+1}}\hspace{-2mm}\de \chi'\frac{W_{i}(\chi)\,n_{j}(\chi')}{D}\,\left\langle\epsilon_a \, \epsilon_a'\right\rangle\Big(\vec{r}(\chi, \chi')\Big)
.
\label{eq:GI2dcorr}
\end{equation}
Once again, $a \in \left\{+,\times\right\}$, where the negative sign is for $a=+$ and the positive one for $a=\times$.

It is quite noteworthy that the GI-alignments are nonzero for $i\neq j$, i.e. while for II-alignments, there are only diagonal entries in the correlation matrix, GIs can occupy all entries. It is obvious from equation \ref{eq:GI2dcorr} that GI-alignments are asymmetric in their nature: Lensing can only happen with structures in front of the sources, therefore there cannot be an IG-correlation.

In order to keep statistical isotropy, however, we apply a symmetrisation and discard the information that in a GI-correlation the more distant galaxy is lensed, which is readily available due to the known galaxy redshift. 
\begin{equation}
C^{\text{GI}}_{a,ij}(\theta) = C^{\text{GI}}_{a,ji}(\theta) = \frac{1}{2} \left\langle\gamma_{a,i} \, \epsilon_{a,j}'\right\rangle.
\end{equation}
Then we can define, like in eq.~\ref{eq:Cpmi},
\begin{equation}
C^{\epsilon,\text{GI}}_{\pm,ij}(\theta) = C^{\text{GI}}_{+,ij}(\theta) \pm C^{\text{GI}}_{\times,ij}(\theta),
\end{equation}
and finally find the angular GI-spectra as
\begin{align}
C^{\epsilon,\text{GI}}_{{E},ij}(\ell)&=\pi \int \theta \, \de \theta ~\Big(C^{\epsilon,\text{GI}}_{+,ij} (\theta) \, J_0(\ell\theta)+C^{\epsilon,\text{GI}}_{-,ij}(\theta)J_4(\ell\theta) \Big),\\
C^{\epsilon,\text{GI}}_{{B},ij}(\ell)&=\pi \int \theta \, \de \theta ~\Big(C^{\epsilon,\text{GI}}_{+,ij} (\theta) \, J_0(\ell\theta)-C^{\epsilon,\text{GI}}_{-,ij}(\theta)J_4(\ell\theta) \Big).
\end{align}
This result implies the presence of $B$-modes from GI-Alignment; note that the contributions of the GI-correlations are mostly negative because of the negative sign in the case of the $\epsilon_+$-component. Apart from these natural GI-contributions, we will treat lensing and IAs independently and ignore all effects of lensing operating on an already intrinsically aligned ellipticity field \citep{giahi-saravani_weak_2014} or evolution of ellipticity correlation functions \citep{giahi_evolution_2013}.

% ---  --- %
\subsection{Alignments of spiral galaxies: tidal torquing-model}\label{sect_torquing}
The alignment of spiral galaxies is thought to be an orientation effect that is related to the direction of the galaxy's angular momentum relative to the line of sight \citep{croft_weak-lensing_2000, crittenden_spin-induced_2001, crittenden_discriminating_2002}. In this picture, intrinsic shape correlations arise because neighbouring galaxies form from correlated initial conditions, which results in correlated angular momentum directions \citep{catelan_evolution_1996, theuns_angular_1997}, in correlated inclination angles under which galactic discs are viewed and consequently in correlated ellipticities \citep{catelan_intrinsic_2001}. Under the assumption that the symmetry axis of the galactic disc coincides with the angular momentum direction $\hat{L}_\alpha=L_\alpha/L$ of the system, one obtains for the ellipticity
\begin{equation}
\epsilon = \frac{\hat{L}_x^2-\hat{L}_y^2}{1+\hat{L}_z^2} +2 \ii \frac{\hat{L}_x\hat{L}_y}{1+\hat{L}_z^2}.
\end{equation}
The angular momentum is generated by tidal gravitational torquing \citep[for a review, see][]{schaefer_review:_2009} of the large-scale structure onto a protogalactic cloud \citep{white_angular_1984, stewart_angular_2013, barnes_angular_1987, theuns_angular_1997} and can effectively be described by a (Gaussian) distribution $p(\hat{L}|\hat{\Phi}_{\alpha\beta})\dd\hat{L}$ for the angular momentum direction $\hat{L}$ with the covariance $\bra\hat{L}_\alpha\hat{L}_\beta\ket$ \citep{lee_galaxy_2001}
\begin{equation}
\bra \hat{L}_\alpha\hat{L}_\beta\ket = 
\frac{1}{3}\left(\frac{1+A}{3}\delta_{\alpha\beta}-A\sum_\gamma\hat{\Phi}_{\alpha\gamma}\hat{\Phi}_{\gamma\beta}\right),
\label{eqn_lmodel}
\end{equation}
that carries the dependence on the orientation of the traceless, unit-normalised tidal shear field $\hat{\Phi}_{\alpha\beta}$ with the properties $\sum_\alpha\hat{\Phi}_{\alpha\alpha}=0$ and $\sum_{\alpha\beta}\hat{\Phi}_{\alpha\beta}\hat{\Phi}_{\beta\alpha}=1$. 

Using the distribution $p(\hat{L}|\hat{\Phi}_{\alpha\beta})\dd\hat{L}$ and the relationship between ellipticity and angular momentum direction one can express the ellipticity in terms of the tidal shear,
\begin{equation}
\epsilon(\hat{\Phi}) = 
\frac{A}{2}\sum_\alpha\left(\hat{\Phi}_{x\alpha}\hat{\Phi}_{\alpha x}-\hat{\Phi}_{y\alpha}\hat{\Phi}_{\alpha y}-2\ii\hat{\Phi}_{x\alpha}\hat{\Phi}_{\alpha y}\right).
\label{eqn_complex}
\end{equation}
and ultimately trace the ellipticity correlations back to a 4-point function of the traceless, unit-normalised tidal shear field $\hat{\Phi}_A$, which in turn can be decomposed into correlations $\tilde{C}_{AB} = \bra\tilde{\Phi}_A\tilde{\Phi}_B\ket$ of the traceless tidal shear $\tilde{\Phi}_A$ field \citep{natarajan_angular_2001,crittenden_spin-induced_2001} by virtue of the Wick-theorem,
\begin{equation}
\bra\hat{\Phi}_A(\bmath{x})\hat{\Phi}_B(\bmath{x})\:\hat{\Phi}_C(\bmath{x}^\prime)\hat{\Phi}^\prime_D(\bmath{x}^\prime)\ket = 
\frac{1}{(14\zeta_2(0))^2}\left(\tilde{C}_{AC}\tilde{C}_{BD}+\tilde{C}_{AD}\tilde{C}_{BC}\right),
\label{eqn_tidal_correlation}
\end{equation}
where capital letters denote index pairs. As in the case of elliptical galaxies, correlations in three dimensions can be Limber-projected for each of the tomography bins and finally Fourier-transformed to yield the corresponding angular ellipticity spectra. We choose a value of $A = 0.25$ for the misalignment parameter $A$ which is found in numerical simulation of tidal torquing \citep{crittenden_spin-induced_2001}. The amplitude of the resulting ellipticity correlations will scale with $A^2$, and will not depend on the normalisation $\sigma_8$ of the fluctuations because as an orientation effect, alignments of spiral galaxies depend only on the unit-normalised tidal shear. A second reason to choose a conservative small value for $A$ is a possible misalignment between the symmetry axis of the galactic disc and the host halo angular momentum direction, as well as a finite thickness of the galactic disc. Thirdly, a structured non-uniform disc will cause a natural scatter in the measured ellipticities even if the disc would be perfectly aligned with the host halo's angular momentum direction. These effects effectively weaken the relationship between shape and angular momentum and decrease the correlation strength, which led us to a conservative choice for $A$ \citep{schaefer_angular_2015, capranico_intrinsic_2013}. We ignore in this model effects of a possible nonlinear evolution of angular momentum \citep[see, e.g.][]{catelan_non-linear_1996,catelan_non-linear_1997, lee_nonlinear_2007}. We would like to point out that in contrast to the alignment model for spiral galaxies constructed by \cite{blazek_beyond_2017}, our model reflects a pure orientation effect and does not depend on the magnitude of tidal gravitational fiels, which is reached by the usage of the unit-normalised tidal shear. Consequently, our alignment model for spiral galaxies does not depend on $\sigma_8$ and only on the shape of the CDM-spectrum through ratios of the spectral moments $\zeta_n(r)$: This will be the reason why we consider $\Omega_\mathrm{m}$ and $h$ in Figs.~\ref{fig:Derivs_omega} and~\ref{fig:Derivs_h}, because they define the shape parameter $\Gamma = \Omega_\mathrm{m} h$ to which the spectral moments are sensitive.

% ---  --- %
\section{Ellipticity correlations}\label{sect_ellipticity}
In this section we show our results angular correlation functions and and angular spectra of the II- and GI-ellipticities for elliptical galaxies in tomographic surveys, and we compare these correlation functions to the II-correlations obtained in the tidal torquing model for spiral galaxies. There is no GI-contribution for spiral galaxies in Gaussian random fields, because the ellipticity correlation function can be traced back to a 3-point function of the tidal shear field, which is naturally zero for Gaussian statistics. For the same reason one would not expect cross-correlations between the intrinsic ellipticities of spiral and elliptical galaxies either.

% ---  --- %
\subsection{Angular ellipticity correlation functions}
\begin{figure}
	\includegraphics[width= \columnwidth]{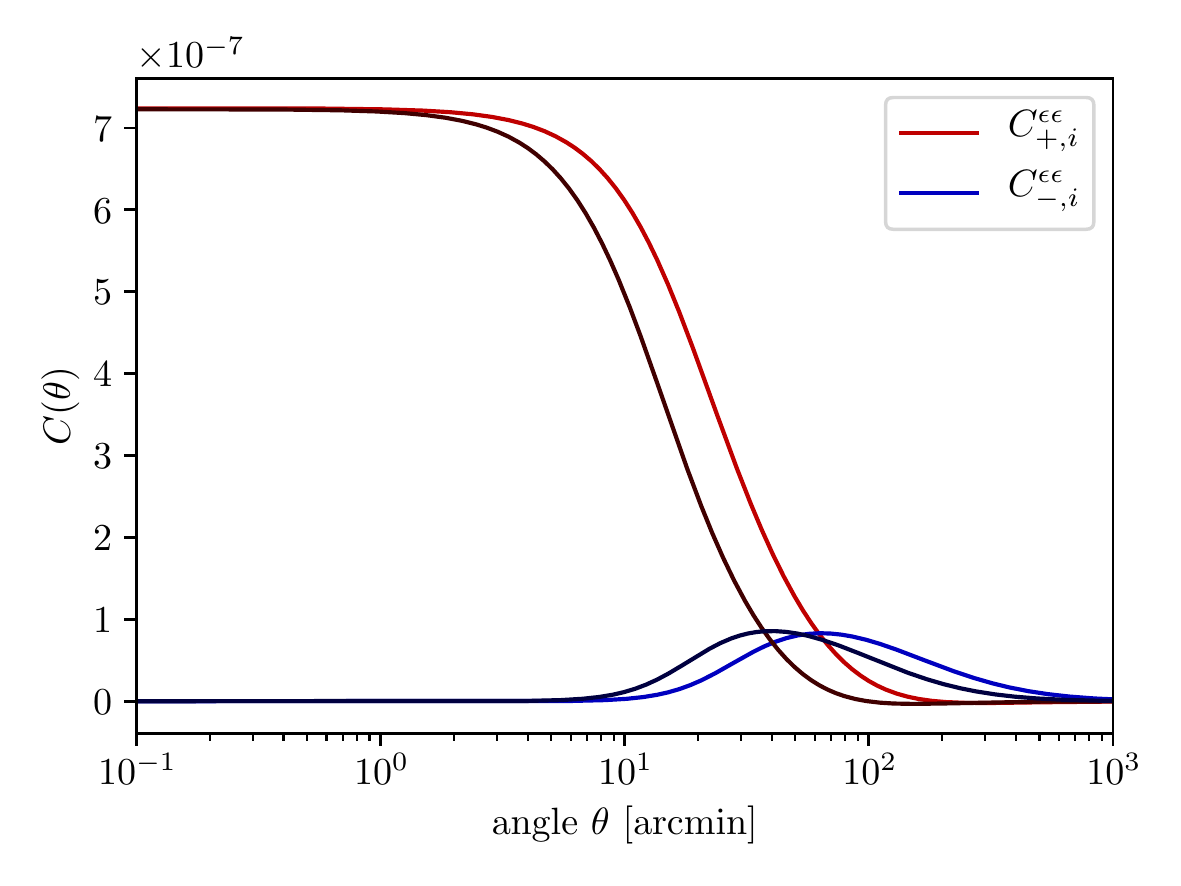}
    \caption{Angular correlation functions $C^{\epsilon \epsilon}_{\pm,i}(\theta)$ for II-correlations in the linear alignment model, for $n_\mathrm{bin}=2$ tomography bins.}
    \label{fig:corr_ii}
\end{figure}

\begin{figure}
	\includegraphics[width= \columnwidth]{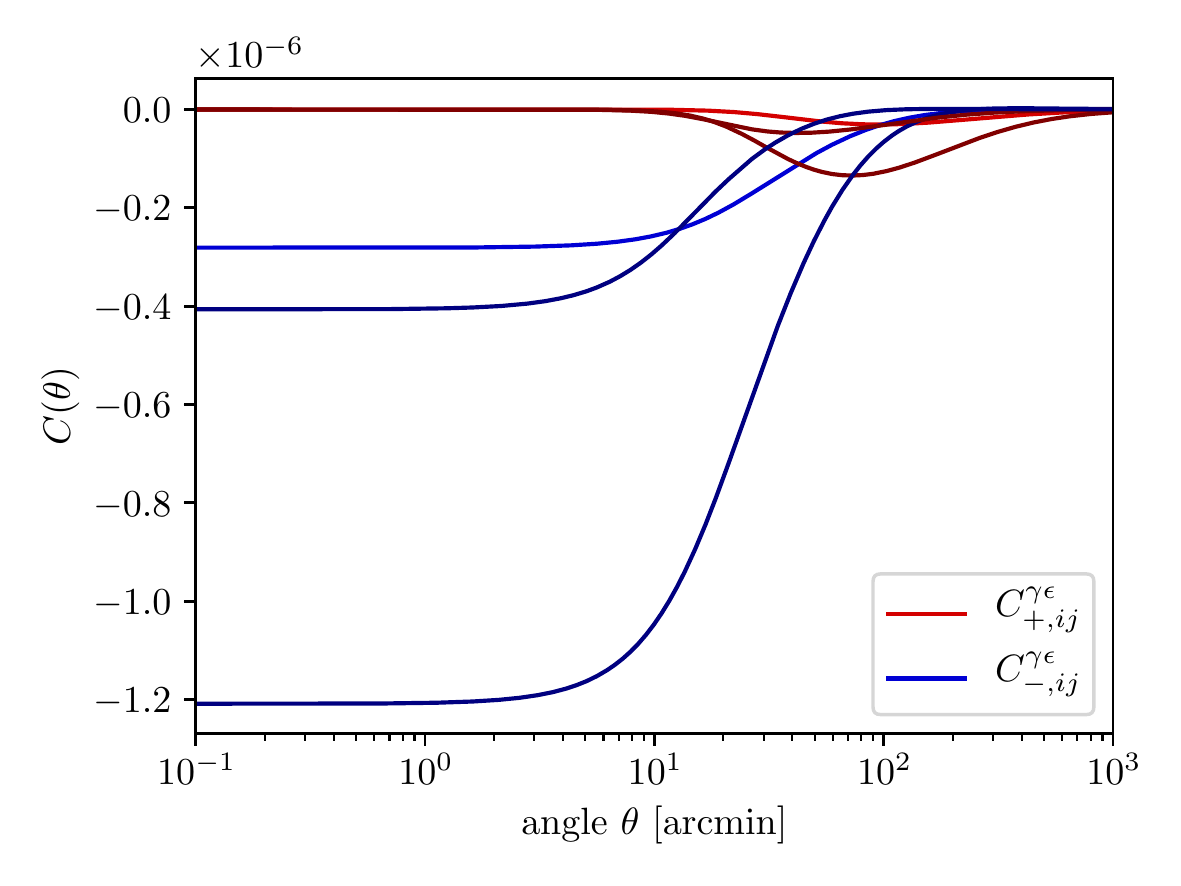}
    \caption{Angular correlation functions $C^{\gamma \epsilon}_{\pm,ij}(\theta)$ for GI-correlations for $n_\mathrm{bin}=2$ tomographic bins.}
    \label{fig:corr_gi}
\end{figure}

The angular ellipticity correlation functions for spiral and elliptical galaxies resulting from the Limber-projection with Euclid's source redshift distribution are shown in Figs.~\ref{fig:corr_ii} and~\ref{fig:corr_gi}. Note that GI-correlations have a negative sign because the gravitational lensing effect is anticorrelated with the physical distortion of an elliptical galaxy in the gravitational field of a lens. However, it is not impossible to construct both positive and negative angular power spectra from them, as $C_+$ and $C_-$ intersect. The II-case produces $n_\mathrm{bin}$ correlation functions, since it is purely local and thus diagonal in the correlation matrix, the GI-case finds $n_\mathrm{bin}(n_\mathrm{bin}+1)/2$ entries, being a non-local cross-correlation. Although there are around a factor of $(n_\mathrm{bin}+1)/2$ more observable spectra for GI-correlations, the overall amplitudes in the II-case are 1-1.5 orders of magnitude larger than for the GI-correlation functions. The linear model, in both cases, produces nonzero correlations up to larger angles $\simeq10^3\,\mathrm{arcmin}$ compared to the quadratic model \citep{schaefer_angular_2015}. Therefore, we expect the corresponding spectra to peak at lower $\ell$ than the ones from the quadratic model.

% ---  --- %
\subsection{Angular ellipticity spectra}
\begin{figure}
	\includegraphics[width=\columnwidth]{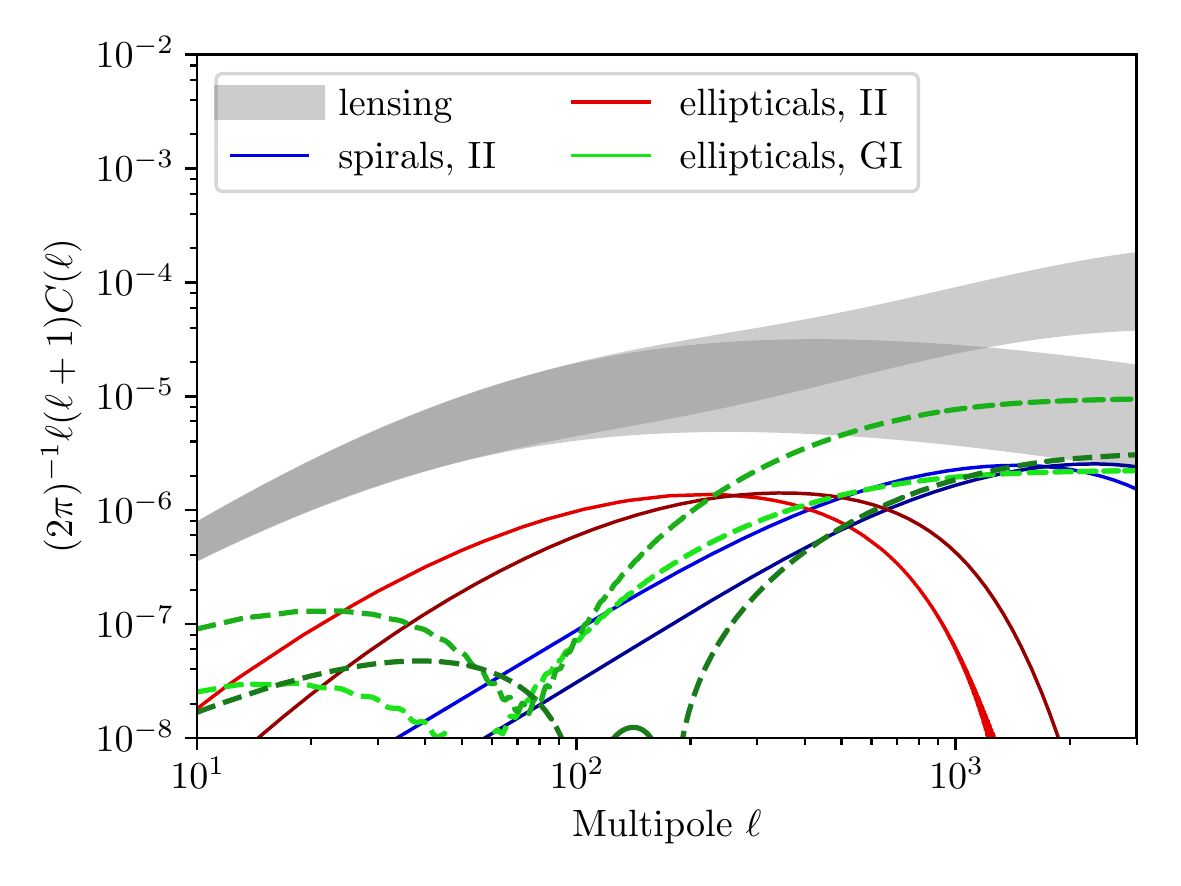}
    \caption{$E$-mode spectra $C^{\epsilon,\mathrm{II}}_{E,ii}(\ell)$, $C^{\epsilon,\mathrm{GI}}_{E,ij}(\ell)$ and $C^\gamma_{E,ij}(\ell)$ for $2$-bin tomography. The grey area represents the linear and non-linear weak lensing contributions, blue shades are II-contributions by spiral galaxies, red shades correspond to II-spectra of elliptical galaxies, green lines are GI-contributions by elliptical galaxies galaxies and the grey bands indicate the amplitudes weak lensing spectra. Dashed lines represent negative contributions.}
    \label{fig:EBspectra_e}
\end{figure}

\begin{figure}
	\includegraphics[width=\columnwidth]{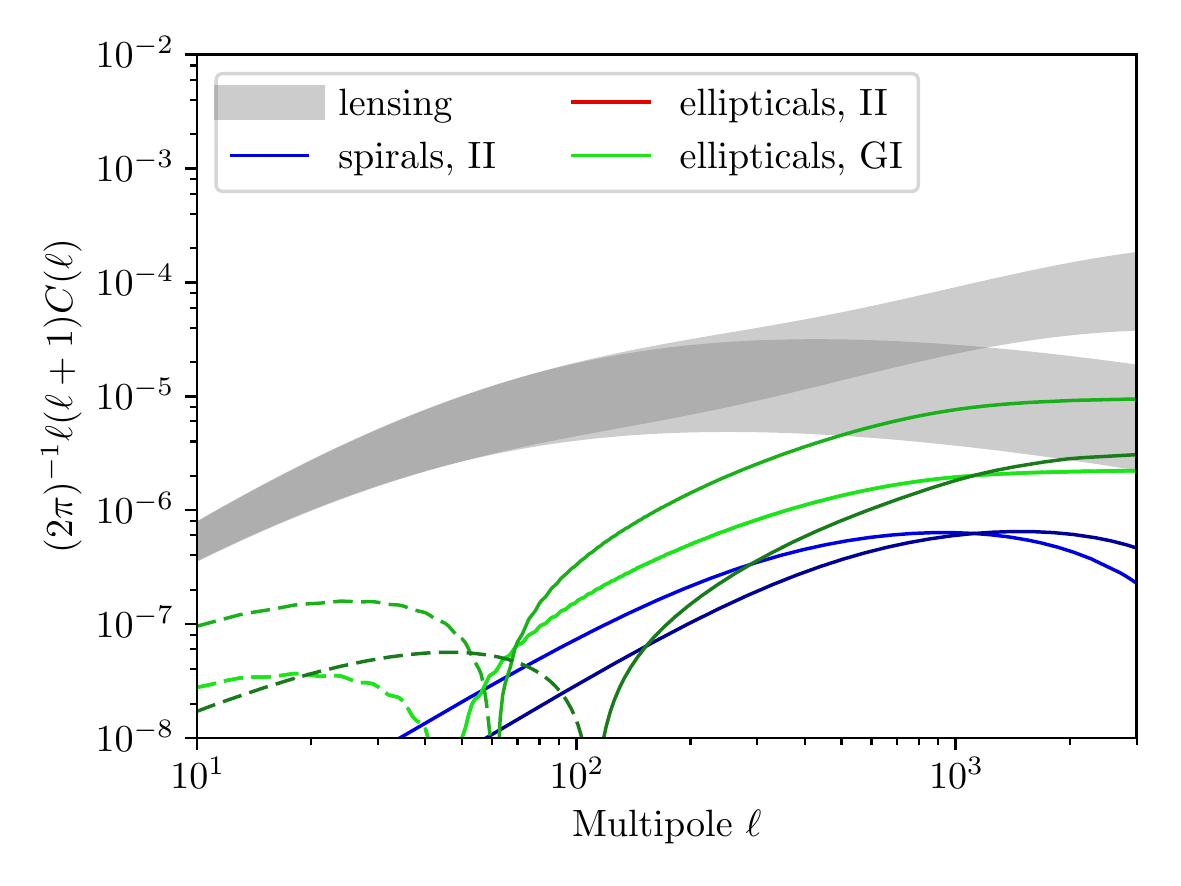}
    \caption{$B$-mode spectra $C^{\epsilon,\mathrm{II}}_{B,ii}(\ell)$ and $C^{\epsilon,\mathrm{GI}}_{B,ij}(\ell)$ for $2$-bin tomography. The grey area represents the linear and non-linear weak lensing contributions, blue shades are II-contributions by spiral galaxies, red shades correspond to II-spectra of elliptical galaxies, green lines are GI-contributions by elliptical galaxies. Dashed lines represent negative contributions, and although weak lensing is $B$-mode free to lowest order, we have included the spectra $C^{\gamma}_{E,ij}(\ell)$ for comparison as the grey bands.}
    \label{fig:EBspectra_b}
\end{figure}

Tomographic angular spectra for the ellipticity field are depicted in Figs.~\ref{fig:EBspectra_e} and~\ref{fig:EBspectra_b} for $2$ tomographic bins. The upper panel shows $E$-mode spectra, the lower panel $B$-modes. As expected with the $E$-modes, the red and green graphs representing elliptical galaxies' II- and GI-cases are larger in amplitude at lower $\ell$ than the spiral contributions in blue. In fact, the elliticals' II-spectrum peaks around $\ell\simeq 300$, with significant contributions of the GI-spectrum at very large ($\ell \simeq 30$) and relatively small ($\ell\simeq 1000$) angular scales. The spirals' spectrum also peaks at small angular scales, around $\ell\simeq 1000$. The ellipticals' GI-contributions are overwhelmingly negative, with very short dips into positive signature at $\ell\simeq 150$, quickly changing back to a negative sign. With increasing the tomographic bin number, this feature shifts a bit in scale ($70 \leq \ell \leq 200$ with 7-bin tomography). We observe that double sign change only happens to off-diagonal spectra ($i\neq j$). At high $\ell$, both the spectra from the spirals and the GI-spectra of the ellipticals start competing in strength with the linear weak lensing signal.

$B$-modes from the linear model are nonexistent in the II-case, although we find a strong GI $B$-mode signal with a sign change for all GI-spectra from negative to positive at intermediate $\ell$. The amplitudes of the GI-signal are generally higher than the ones for the spiral $B$-mode signal, its signal strength for high $\ell$ comparable to the one of the linear weak lensing spectrum. 

\begin{figure*}
	\includegraphics[width=2 \columnwidth]{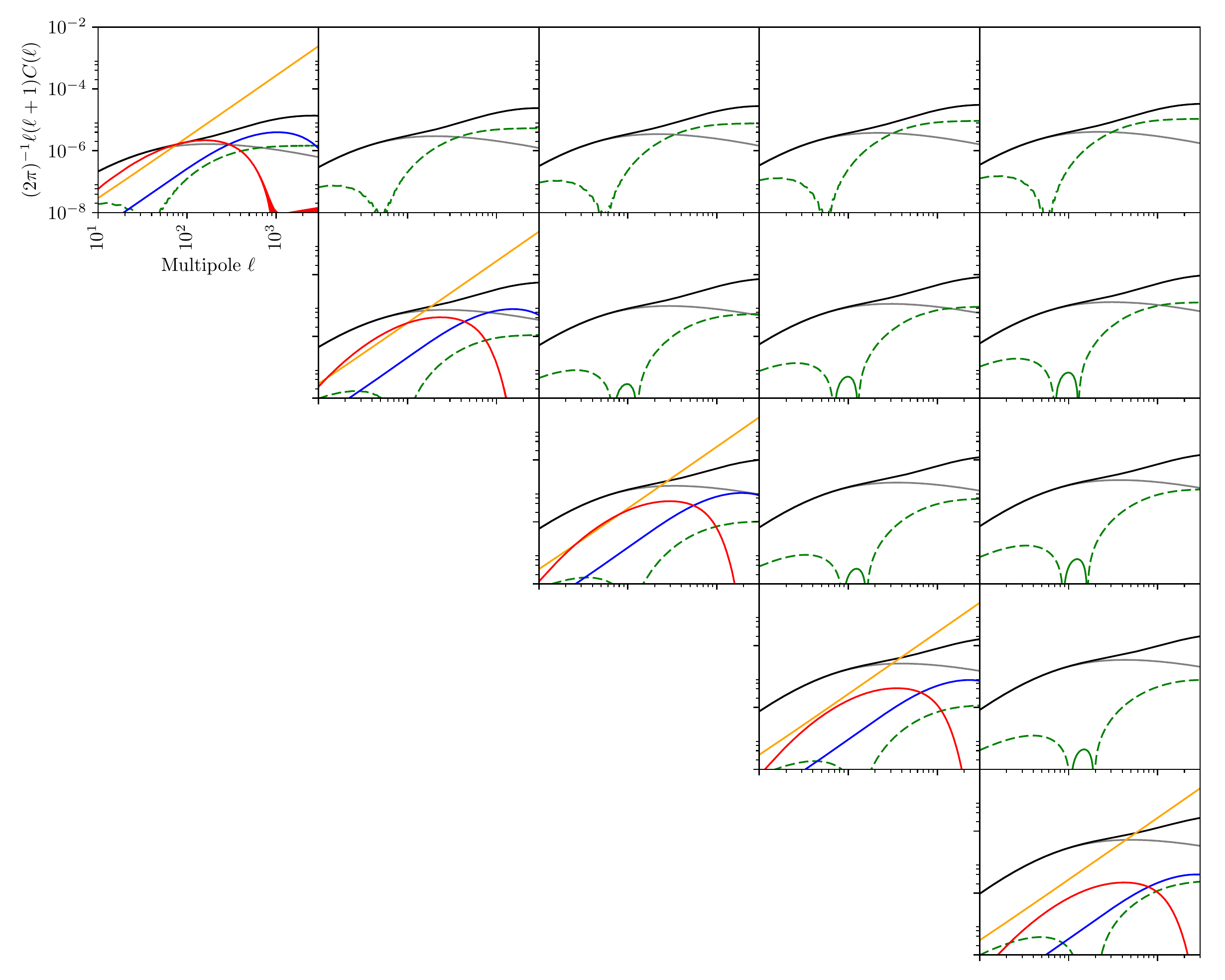}\\
	\caption{$E$-mode spectra $C^{\gamma}_{E,ij}(\ell)$, $C^{\epsilon,\mathrm{GI}}_{E,ij}(\ell)$ and $C^{\epsilon,\mathrm{II}}_{E,ii}(\ell)$ for $5$-bin tomography; the bin correlations are ordered matrix-like from top left ($C_{E,11}(\ell)$) to bottom right ($C_{E,55}(\ell)$, with increasing redshift from bin 1 to bin 5). Black curves show the weak lensing signal using a nonlinear CDM-spectrum $P(k)$, while the grey curves use a linear CDM-spectrum. Ellipticity spectra of the II-type are shown in blue for spiral galaxies and in red for elliptical galaxies. The GI-spectra for elliptical galaxies are given in  green, where solid lines indicate positive and dashed lines negative values. The orange lines depict the ellipticity shape noise $\sigma_\epsilon^2 n_\mathrm{bin}/\bar{n}$ for a tomographic measurement with $n_\mathrm{bin} = 5$ redshift bins.}
	\label{fig:spectramatrix}
\end{figure*}

All tomographic spectra for a 5-bin application are shown by Fig.~\ref{fig:spectramatrix}: Clearly one sees the trend that lensing dominates over intrinsic alignments at high redshifts for both spiral and elliptical galaxies, which are predicted by the tidal torquing and tidal shearing models, respectively, adding weight to the proposal by \citet{kirk_optimising_2011} to use this property for survey optimisation. Interestingly, this strategy would be applicable even to the case of GI-alignments, and it shows that in inference from the off-diagonal spectra $C^{\gamma}_{E,ij}(\ell)$ and $C^{\epsilon,\mathrm{GI}}_{E,ij}(\ell)$, $i\neq j$, as proposed by \citet{king_separating_2003}, only the GI-signal would need to be taken care of. 

% ---  --- %
\section{Statistical errors}\label{sect_statistical}

% ---  --- %
\subsection{Amplitude of the intrinsic ellipticity spectrum}
The first question concerns the amplitude of the different contributions to the ellipticity spectrum. While weak lensing is computed straightforwardly in a parameter-free way, this is not the case for intrinsic alignments, whose amplitude depends on the details of the tidal interaction processes. Specifically, in the case of spiral galaxies the amplitude of the alignment signal is set by parameterising the relationship between the orientation of the tidal gravitational field and the angular momentum direction, and by the thickness of the galactic disc. These parameters are roughly known from $n$-body simulation of structure formation and from the observation of galaxies in the local universe, respectively. For elliptical galaxies, one needs information about how strongly the stellar component of an elliptical galaxy reacts to tidal gravitational fields by changing its shape: While in principle this parameter can be measured from simulations as well, we derive an estimate of this parameter from the nonzero alignment signal for elliptical galaxies measured by CFHTLenS\footnote{http://www.cfhtlens.org}.

In fact, CFHTLenS has measured a significant alignment signal for elliptical galaxies, and we adjust our model parameter, the constant of proportionality between tidal shear field and observed ellipticity, to reproduce the CFHTLenS-signal \citep{Heymans:2013fya}: To this purpose, we adopted its survey characteristics (galaxy density $n=11/\mathrm{arcmin}^2$, $\beta=3/2$, $z_\mathrm{med}=0.7$ and $f_\mathrm{sky}=37\times10^{-4}$) and set up a computation of the signal to noise-ratio $\Sigma$ that one can obtain for a measurement of the intrinsic contribution of the ellipticity spectrum of elliptical galaxies in a tomographic survey with the CFHTLenS redshift bins.

The signal to noise-ratio $\Sigma = S_0/\sigma_S$ is given by the ratio between the unknown signal amplitude $S_0$ in units of the statistical error $\sigma_S^2$ on measuring $S_0$,
\begin{equation}
\Sigma^2 = f_\mathrm{sky}\sum_\ell\frac{2\ell+1}{2}\mathrm{tr}\left(C^{-1}S\right)^2,
\end{equation}
where we assume Gaussian statistics for estimating the statistical error $\sigma_S$ using the Fisher-formalism,
\begin{equation}
\frac{1}{\sigma_S^2} = f_\mathrm{sky}\sum_\ell\frac{2\ell+1}{2}\mathrm{tr}\left(C^{-1}\partial_{S_0}C\right)^2,
\end{equation}
where naturally the spectrum $C(\ell) = S(\ell) + N(\ell)$ decomposes into a signal part $S(\ell)$ that describes the ellipticity correlations of elliptical galaxies and the cross-correlations between intrinsic alignments and weak lensing, as well as a noise part $N(\ell)$. Here, only the signal $S(\ell)$ is proportional to powers of $S_0$, while the noise $N(\ell)$ contains weak lensing, the intrinsic alignments of spiral galaxies and the shape noise contribution in addition. Converting into the alignment parameter $D$, we find $D = 9.5\times10^{-5}\,c^2$ to reproduce the CFHTLenS-observation, which is comparable to results from numerical simulations \citep[e.g.][who report $D = 1.5\times10^{-4}c^2$ in our choice of units,]{hilbert_intrinsic_2016}. At the same time we have verified that the alignments of spiral galaxies would not yield a measurable signal for CFHTLenS, so we continue to use parameters that have been obtained from numerical simulations and from the ellipticity distribution of galaxies in the local Universe.

% ---  --- %
\subsection{Estimation of statistical errors}
In the extraction of cosmological information from weak lensing data one introduces the likelihood $\likeli\left(\left\{\gamma_{\ell m,i}\right\}\right)$, which is the probability to obtain a set of spherical harmonics expansion coefficients $\left\{\gamma_{\ell m,i}\right\}$ in the model $\hat{C}^\gamma_{E,ij}(\ell)$,
\begin{equation}
\likeli\left(\left\{\gamma_{\ell m,i}\right\}\right) = 
\prod_\ell \likeli\left(\gamma_{\ell m,i}|\hat{C}^\gamma_{E,ij}(\ell)\right)^{2\ell+1},
\end{equation}
where we will consider only the $E$-modes of the observed ellipticity field as a source of cosmological information. Due to the assumed full sky-coverage and statistical isotropy the likelihood factorises in $2\ell+1$ statistically equivalent $\gamma_{\ell m,i}$-modes for each tomographic bin $i$. These modes are obtained by transforming the shear field $\gamma_i(\bmath\theta)$
\begin{equation}
\gamma_{\ell m,i} = \int\dd\Omega\: \gamma_i(\bmath\theta)Y_{\ell m}^*(\bmath\theta).
\end{equation}
Under the assumption that the modes $\gamma_{\ell m,i}$ are Gaussian distributed with zero mean, the likelihood takes the form
\begin{equation}
\likeli\left(\gamma_{\ell m,i}\right) = 
\frac{1}{\sqrt{(2\pi)^{n_\mathrm{bin}}\mathrm{det}\hat{C}^\gamma_{E,ij}(\ell)}}\exp\left(-\frac{1}{2}\gamma_{\ell m,i}(\hat{C}^\gamma_{E}(\ell)^{-1})_{ij}\gamma_{\ell m,j}\right).
\end{equation}
We will work for simplicity under the Gaussian assumption, although there should be an impact from non-Gaussianity of the observable on the likelihood and therefore on the parameter estimation process \citep{scoccimarro_power_1999, scoccimarro_fitting_2001, van_waerbeke_weak_2001, casarini_tomographic_2012, kayo_information_2013}. For scales that are nonlinearly evolving, we increase the variance by using a nonlinear CDM-spectrum $P_\delta(k)$ but keep a Gaussian form for the distribution of modes.

The logarithmic likelihood $L = -\ln\likeli$
\begin{equation}
L = 
\sum_\ell\frac{2\ell+1}{2}\:
\left(\trace\:\ln\:\hat{C}^\gamma_{E,ij}(\ell) + \left(\hat{C}^\gamma_{E}(\ell)^{-1}\right)_{ij}\:\gamma_{\ell m,i}\gamma_{\ell m,j}\right),
\end{equation}
can then be used to define the Fisher-matrix $F_{\mu\nu} = \bra\partial_\mu\partial_\nu L\ket$ as the data-averaged negative curvature at the location of the highest likelihood, i.e. the fiducial cosmology:
\begin{equation}
F_{\mu\nu} = \sum_\ell\frac{2\ell+1}{2}
\trace
\left(
\partial_\mu\ln\hat{C}^\gamma_{E,ij}(\ell)\:\partial_\nu\ln\hat{C}^\gamma_{E,jk}(\ell)\right).
\label{eqn_fisher}
\end{equation}
The partial derivatives are taken with respect to the cosmological parameter set, $\left\{\Omega_\mathrm{m},\sigma_8,h,n_s,w\right\}$.

The characteristics of Euclid's weak lensing survey \citep{laureijs_euclid_2011} are summarised as $(i)$ a median redshift of $0.9$, $(ii)$ a yield of $\bar{n}=4.7\times10^8$ galaxies per unit solid angle, $(iii)$ a sky fraction of $f_\mathrm{sky}\simeq 0.363$ and $(iv)$ a Gaussian shape noise with standard deviation $\sigma_\epsilon=0.4$. We work with modes $\gamma_{\ell m,i}$ that are statistically independent in the wave numbers $\ell$ and $m$ despite incomplete sky coverage implying that the likelihood would not separate, but instead scale the logarithmic likelihood with a factor $\sqrt{f_\mathrm{sky}}$. The tomographic bins are chosen to contain the same fraction $1/n_\mathrm{bin}$ of galaxies. We always consider the $E$-mode of the observed ellipticity field as the source of cosmological information.

\begin{figure}
	\includegraphics[width=\columnwidth]{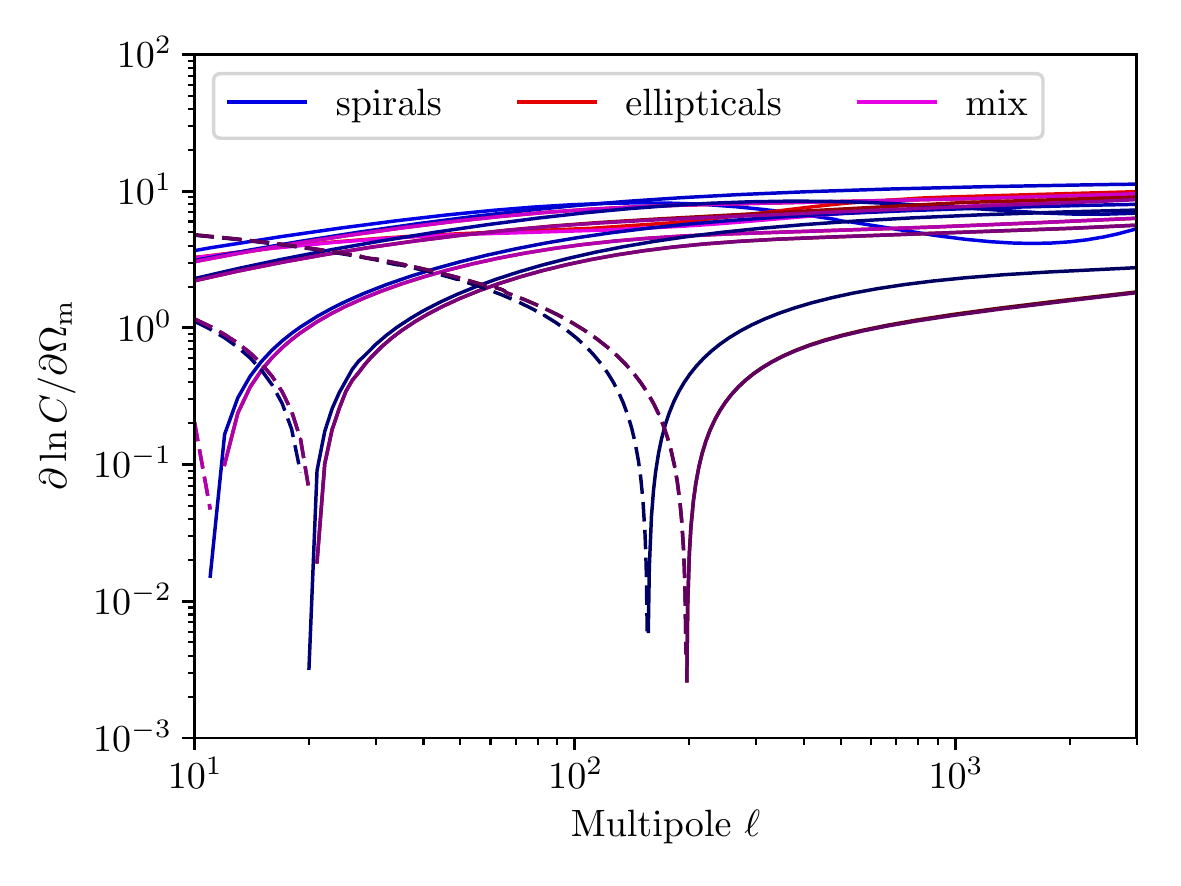}

    \caption{Logarithmic derivatives of the $E$-mode spectra $C^{\epsilon,\mathrm{II}}_{E,ii}(\ell)$ with respect to $\Om$. Blue shades correspond spiral galaxies, red shades to elliptical galaxies, purple considers a combined spectrum weighed with the spiral fraction $q=0.7$. Dashed lines are negative contributions.}
    \label{fig:Derivs_omega}
\end{figure}

\begin{figure}
	\includegraphics[width=\columnwidth]{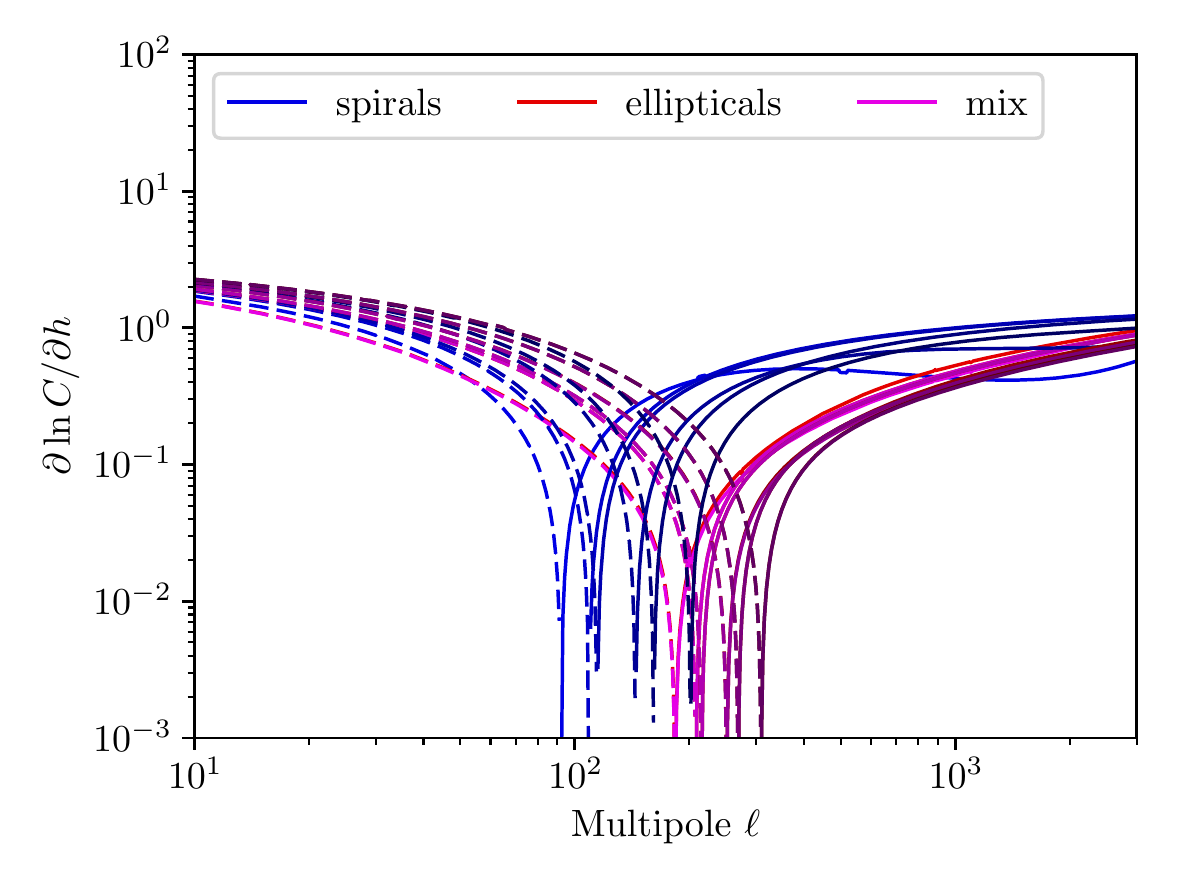}

    \caption{Logarithmic derivatives of the $E$-mode spectra $C^{\epsilon,\mathrm{II}}_{E,ii}(\ell)$ with respect to $h$. Blue shades correspond to only spiral galaxies, red shades to elliptical galaxies, purple considers a combined spectrum weighed with the spiral fraction $q=0.7$. Dashed lines are negative contributions.}
    \label{fig:Derivs_h}
\end{figure}

Figs.~\ref{fig:Derivs_omega} and~\ref{fig:Derivs_h} show logarithmic derivatives $\partial_\mu\ln C$ of the intrinsic alignment covariances with respect to cosmological parameters $\Om$ and $h$, split up into elliptical and spiral galaxies assuming a spiral fraction of $q=0.7$. These logarithmic derivatives in addition to those from weak lensing would enter the construction of the Fisher-matrix eq.~\ref{eqn_fisher}. From the Fisher-matrix we derive Cram{\'e}r-Rao bounds $\sigma^{-2}_{\mu\mu}\leq F_{\mu\mu}$ and marginalised likelihoods for comparison with the systematic errors. Both cosmological parameters affect the intrinsic alignment spectrum through the shape of the CDM-spectrum. In addition, $\Om$ sets the absolute strength of the alignment of elliptical galaxies as it appears as a coupling constant in the comoving Poisson-equation.

We woudl like to point out that there are subtle differences in the parameter sensitivity of the tidal shearing and tidal torquing models \citep{capranico_intrinsic_2013, schaefer_angular_2015, merkel_parameter_2017}, because tidal torquing does not depend on the absolute magnitude of the tidal gravitational fields whereas the tidal shearing would, although both models ultimately express tidal field correlations in terms of the spectral functions $\zeta_n(r)$: The crucial step is the usage of the unit-normalised tidal shear for the tidal torquing model in eq.~\ref{eqn_tidal_correlation}.

% ---  --- %
\section{Systematical errrors}\label{sect_systematical}

% ---  --- %
\subsection{Parameter estimation biases}
Intrinsic alignments are a systematic in weak lensing measurements. If they are present in ellipticity spectra as taken by weak lensing surveys, they would lead to parameter estimation biases if they are not included in the modelling or mitigated in a statistical way. In this section we intend to compute the effect of a composite alignment model on the parameter estimation process, which is particularly important because the inclusion of both elliptical and spiral galaxies generates an intrinsic alignment contribution over a wide range in angular scales, and because the presence of negative GI-cross spectra between the alignments of elliptical galaxies and weak lensing gives rise to a further modulation of the spectrum on intermediate scales. In general it is the case that alignments of elliptical galaxies contribute on multipoles between $\ell=30$ and $\ell=300$, on a level of 1\% of the weak lensing signal, whereas spiral galaxies would generate a similar signal on multipoles above $\ell=300$. GI-alignments of elliptical galaxies dominate on large angular scales and remove amplitude from the ellipticity spectra due to their negative sign.

We consider the specific case of inference of the full parameter set of a $w$CDM-cosmology consisting of $\Omega_\mathrm{m}$, $\sigma_8$, $h$, $n_s$, and $w$ from the tomographic $E$-mode spectrum of the observed ellipticity field $\epsilon_{\mathrm{obs},i}$ in the redshift bin $i$. We define the covariance matrix of the true model $C_t(\ell)$ containing contributions from both lensing and intrinsic alignments and the corresponding covariance matrix $C_f(\ell)$ of the false model wrongly only contains gravitational lensing,
\begin{align}
C_{t,ij}(\ell) & 
= C^{\epsilon,\mathrm{II}}_{E,ij}(\ell)\delta_{ij} + C^{\epsilon,\mathrm{GI}}_{E,ij}(\ell) + C^{\gamma}_{E,ij}(\ell)\\
C_{f,ij}(\ell) & 
= C^{\gamma}_{E,ij}(\ell)
\end{align}
where no summation over the double index $i$ is implied. A fit of of the incomplete model $C_{f,ij}(\ell)$ to data that would follow the true model $C_{t,ij}(\ell)$ would then result in parameter estimation biases, i.e. systematical errors $\delta_\mu$ which are caused by the incompleteness of the model because it does not contain intrinsic alignments. The estimation biases $\delta_\mu$ can be obtained from solving the linear equation
\begin{equation}
\sum_\nu G_{\mu\nu}\delta_\nu = a_\mu\rightarrow \delta_\mu = \sum_\nu (G^{-1})_{\mu\nu}a_\nu,
\end{equation}
which is defined by with the vector $a_\mu$,
\begin{equation}
a_\mu = \sum_\ell\frac{2\ell+1}{2}
\trace\left[\partial_\mu\ln C_f\left(\id-C_f^{-1}C_t\right)\right],
\end{equation}
and the matrix $G_{\mu\nu}$,
\begin{align}
G_{\mu\nu}
& = \sum_\ell\frac{2\ell+1}{2}
\trace
\left[C_f^{-1}\:\partial^2_{\mu\nu}C_f\:\left(C_f^{-1}C_t-\id\right)\right]
\nonumber\\
& - \sum_\ell\frac{2\ell+1}{2}
\trace
\left[\partial_\mu\ln C_f\partial_\nu\ln C_f\:\left(2C_f^{-1}C_t-\id\right)\right],
\end{align}
that need to be computed from the matrices $C_{t,ij}(\ell)$ and $C_{f,ij}(\ell)$ and its derivatives. In the case of choosing the correct model, $C_{t,ij}(\ell) = C_{f,ij}(\ell)$ and consequently $a_\mu=0$, such that no biases arise. $\id$ refers to the unit matrix in $n_\mathrm{bin}$ dimensions and the traces are taken over the tomography bin indices. This formalism, described in \citet{schafer_weak_2012} is a generalisation of the method proposed by \citet{taburet_biases_2009} for correlated data points and improves on \citet{amara_systematic_2008} by taking care of the change in the parameter covariance as a function of model parameters.

\begin{figure*}
\includegraphics[width=2 \columnwidth]{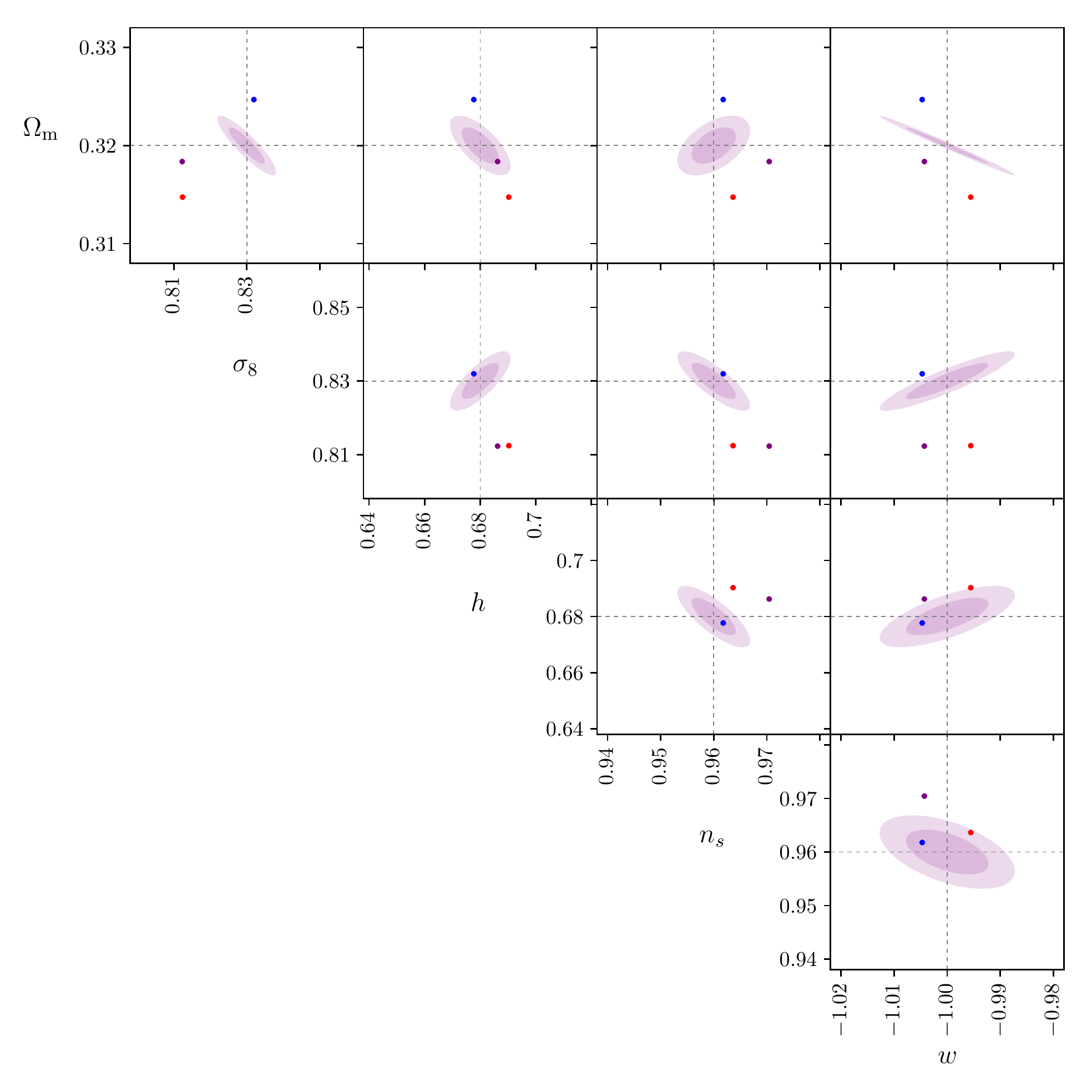}
\caption{Parameter estimation biases for a EUCLID-like survey with $7$ tomographic bins for the cosmological parameters $\Omega_\mathrm{m}$, $\sigma_8$, $h$, $n_s$ and $w$ used in this work. The inner ellipses show $1\sigma$, the outer ellipses $2\sigma$ Fisher contours. Blue dots show the location of the fiducial values with only spiral galaxies in the survey, red dots show the fiducial values with only elliptical galaxies in the survey. Purple dots correspond to a survey with both types weighed by the spiral fraction $q=0.7$.}
\label{fig:biases}
\end{figure*}

As we compute the parameter estimation biases for the full parameter set of a dark energy cosmology in order to simulate the statistical inference process from actual data, we are in a position to quantify which parameters are most strongly affected by the presence of a systematic: In our case the parameters $\Omega_\mathrm{m}$ and $\sigma_8$ would be most strongly biased because they are at the same time determining the amplitude of the weak lensing spectrum to leading order. The systematical error would be, depending on the galaxy type and the number of tomographical bins used, up to twice or three times as large as the statistical error in case of $\Omega_\mathrm{m}$ and $\sigma_8$ respectively. In recent works such as \cite{2016arXiv161004606J}, there is tension between the $\Omega_\mathrm{m}$ and $\sigma_8$ measurements of weak lensing surveys (KiDS\footnote{\url{http://kids.strw.leidenuniv.nl}}) and CMB measurements (Planck\footnote{\url{http://www.esa.int/Our_Activities/Space_Science/Planck}}). In particular, weak lensing systematically underestimates both parameters, which qualitatively supports the direction of bias we found for the mixed model and ellipticals, cf. \reff{fig:biases} and \reff{fig:Fisher_BE}.

Here, a point needs to be made on the linearity of the biases; as seen in \reff{fig:biases}, the red dots (bias only due to elliptical galaxies) and the blue dots (only spiral galaxies) will not, in general, additively give the same results the purple dots (mixture of both models). This is due to the fact that the derivatives of the three different (true) spectra
\begin{align}
C^{\text{red}}_{t,ij}(\ell) & 
= C^{\epsilon,\mathrm{II, e}}_{E,ij}(\ell)\delta_{ij} + C^{\epsilon,\mathrm{GI}}_{E,ij}(\ell) + C^{\gamma}_{E,ij}(\ell),\\
C^{\text{blue}}_{t,ij}(\ell) & 
= C^{\epsilon,\mathrm{II, s}}_{E,ij}(\ell)\delta_{ij} + C^{\gamma}_{E,ij}(\ell),\\
C^{\text{purple}}_{t,ij}(\ell) & 
= C^{\epsilon,\mathrm{II, s}}_{E,ij}(\ell)\delta_{ij} + C^{\epsilon,\mathrm{II, e}}_{E,ij}(\ell)\delta_{ij} + C^{\epsilon,\mathrm{GI}}_{E,ij}(\ell) + C^{\gamma}_{E,ij}(\ell),
\end{align}
are, in general, not additive. This can be seen in e.g. \reff{fig:Derivs_omega}. Here, in particular, two cases exhibit a sign change, whilst the third does not. Such a behaviour makes it obvious that a mixture of both models will not be easily analytically predictable from the individual properties of each model.

The relative weights of the two models, expressed by a constant $q = 0.7$ in our work, is affecting the resulting bias of a mixture of both models as well. A greatly varying $q$ over cosmic time would have to be modelled, however as $q$ has been shown not to vary greatly at least up to medium redshifts and since even future surveys won't go much beyond $z\sim2$, where observational selection effects dominate over cosmic morphology abundances, we assume it safe to fix $q$ to a constant. Additionally, due to the lensing efficiency function, the relative contribution of IAs at high redshift become smaller. Clearly, an extreme case of $q\rightarrow{1,0}$ would reduce to just one model describing the alignment effects, but as long as $q$ stays moderately constant, there are many larger uncertainties in the observations than fluctuations in relative abundances of galaxy types and numerical tests have shown that changing $q$ by 10 per cent won't affect the resulting bias of the mixed model greatly.

\begin{figure}
\includegraphics[width=\columnwidth]{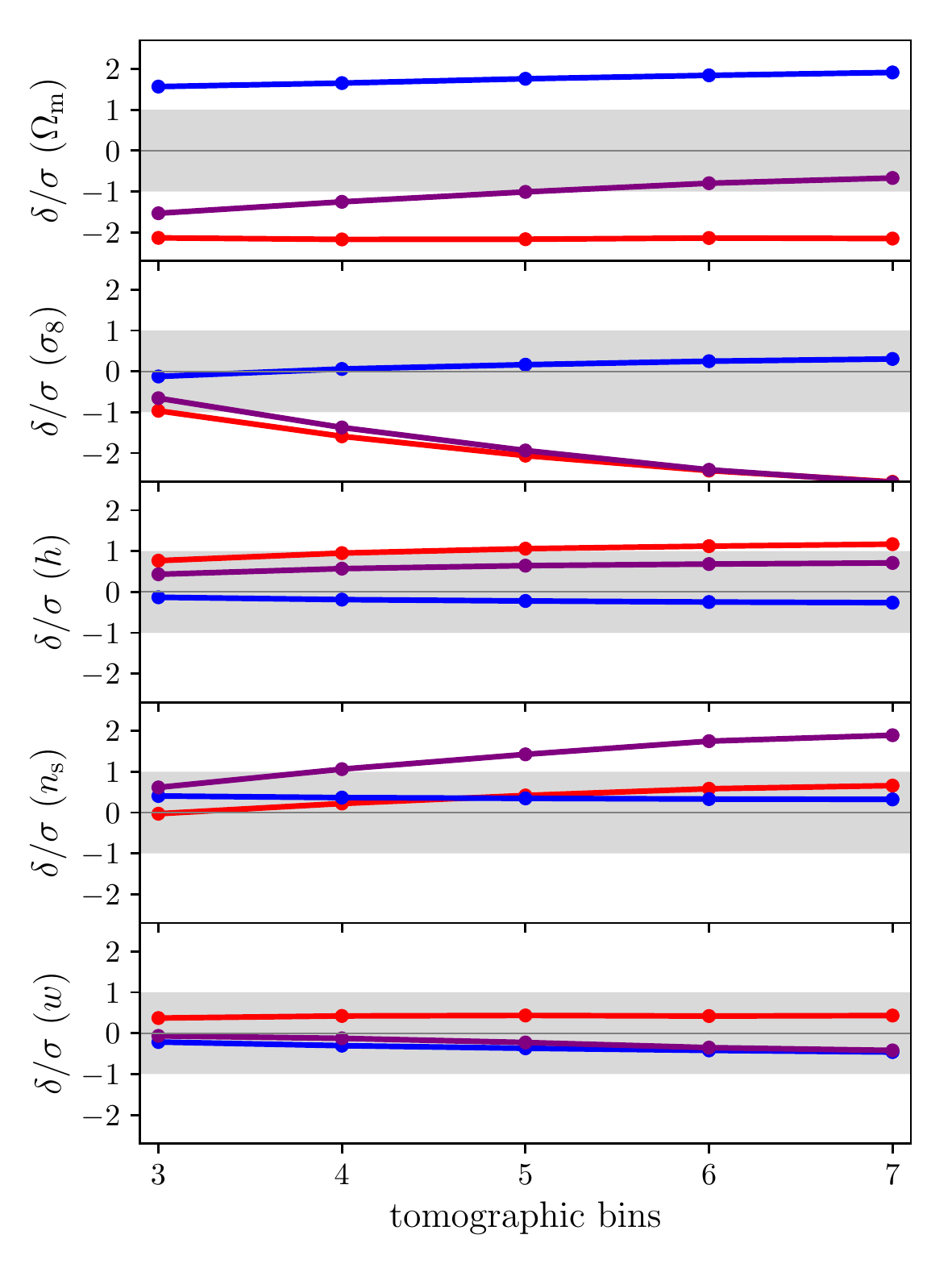}
\caption{Marginalised parameter estimation biases for the same survey setup as in Fig.~\ref{fig:biases} for increasing tomographic bin number $n_\mathrm{bin}$. The bias $\delta_\mu$ for each parameter is normalised by its variance $\sigma^2 = (F^{-1})_{\mu \mu}$, the grey areas mark 1 $\sigma$ around the concordance value.}
\label{fig:MargBiases}
\end{figure}

The dark energy equation of state parameter $w$ is only relatively weakly affected and fall well within the statistical uncertainty. For that reason it would seem implausible that a dynamic dark energy model would be favoured over a cosmological constant $\Lambda$. The situation is less clear for $n_s$, which would be measured too close to unity for all three models, which might lead to wrong inference on the inflationary slow-roll parameters, as the bias can reach up to twice the statistical error. The biases on $h$ are highly significant (i.e. they are on the same order of magnitude as the statistical error) and might in the future be able to contribute to the discussion concerning values of $h$ from different measurements methods.
Biases in units of the statistical error, $\delta_\mu / \sigma_\mu$ show a slow scaling with the number of tomographic bins $n_\mathrm{bins}$ for dividing the galaxy sample, as shown by Fig.~\ref{fig:MargBiases}: With larger $n_\mathrm{bin}$ one usually observes shrinking statistical errors and increasing systematical ones, so the ratio $\delta_\mu / \sigma_\mu$ is usually increasing in magnitude. The strongest overall increase can be seen in the parameters $\sigma_8$ and $n_s$, whereas other parameters show only a weak scaling. In particular, the difference of sensitivity of the two models on $\sigma_8$ reflects their set-up; whereas the observed ellipticity of spiral galaxies induced by tidal torquing is only sensitive to the direction of the shear field $\hat{\Phi}_{\alpha \beta}$ (cf. eq.~\ref{eqn_complex}), the same quantity for elliptical galaxies is sensitive to the shear fields' magnitude as well as direction, i.e. ${\Phi}_{\alpha \beta}$ (see eq.~\ref{eq:linearmodel}). In other words, tidal torque alignment is unaffected the magnitude of angular momentum (induced by shearing), whereas tidal shear alignment scales with the strength of the shear. Hence, one would expect greater sensitivity of the latter towards the amplitude of the gravitational field, which is linked to the amplitude of the fluctuations by the Poisson equation and thus a greater impact on $\sigma_8$. Note that the sensitivity of the tidal shear model towards the magnitude of the gravitational potential can, given sufficient precision in observation and sufficient knowledge about other bias effects, be used as a probe for the local Newtonian potential.

% ---  --- %
\subsection{Loss of Bayesian evidence for $\Lambda$CDM due to systematics}
\begin{figure}
\includegraphics[width = \columnwidth]{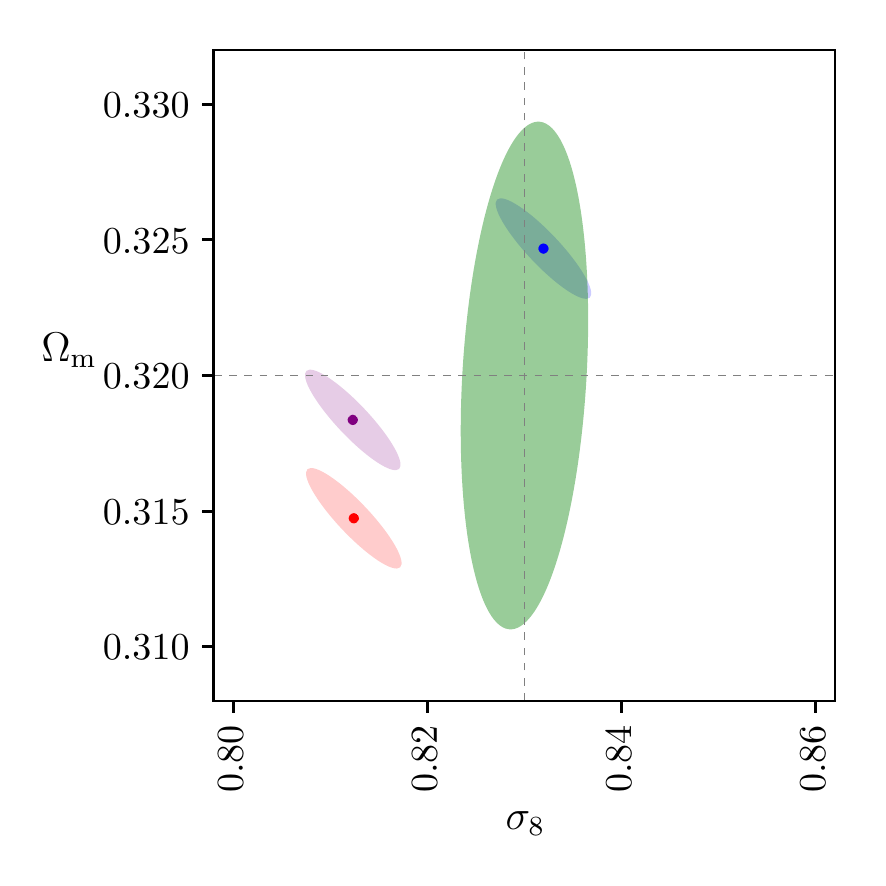}
\caption{Lensing likelihoods $\likeli_f$ in the $\Omega_\mathrm{m}$-$\,\sigma_8$-plane for $(i)$ only ellipticals (red), $(ii)$ only spirals (blue), and $(iii)$ a mix of both (purple), all for $7$ tomographic bins. Their offset from the centre (where the concordance values of $\Omega_\mathrm{m}=0.323 $ and $\sigma_8=0.834$ lie) corresponds the induced systematical error by ignoring intrinsic alignment signals from the respective models. In green, a  CMB prior from Planck is pictured, also centered on the concordance values. All contours are 1$\sigma$.}
\label{fig:Fisher_BE}
\end{figure}

After having obtained systematical and statistical errors we would like to answer the question if the presence of parameter estimation biases can decrease the Bayesian evidence for a specific $\Lambda$CDM parameter set as the reference cosmological model \citep{trotta_applications_2007, trotta_bayes_2008, trotta_bayesian_2017, trotta_forecasting_2007}. If prior information $p(\theta_\mu)$ on the cosmological model centered on concordance values $\theta_{\mu, t}$ is available, for instance from CMB-observations or from galaxy clustering, the presence of systematics specific to weak lensing such as intrinsic alignments, would shift the (true) lensing likelihood $\likeli_t(\theta_\mu)$ relative to the true cosmological model, and the likelihood would become inconsistent with the prior or other cosmological measurements, see Fig.~\ref{fig:Fisher_BE}.

For this purpose, we quantify the loss of Bayesian evidence for $\Lambda$CDM that is caused in a weak lensing survey by the presence of intrinsic alignments, and use Gaussian priors on the cosmological parameter set from CMB temperature and polarisation spectra and from galaxy clustering, for which we use the specifications of Planck and of Euclid, respectively.
We can write the Bayes factor $k$ in terms of the prior $p(\theta_\mu)$, the unbiased likelihood $\likeli_t(\theta_\mu)$, and the biased (i.e. shifted) likelihood $\likeli_f(\theta_\mu)$,
\begin{equation}
k = \frac{\int \dd^N\theta\, p(\theta_\mu)\,\likeli_t(\theta_\mu)}
			{\int \dd^N\theta\, p(\theta_\mu)\,\likeli_f(\theta_\mu)},
\label{eq:BE}
\end{equation}
for an $N$-dimensional parameter space. 

\begin{figure}
\includegraphics[width = \columnwidth]{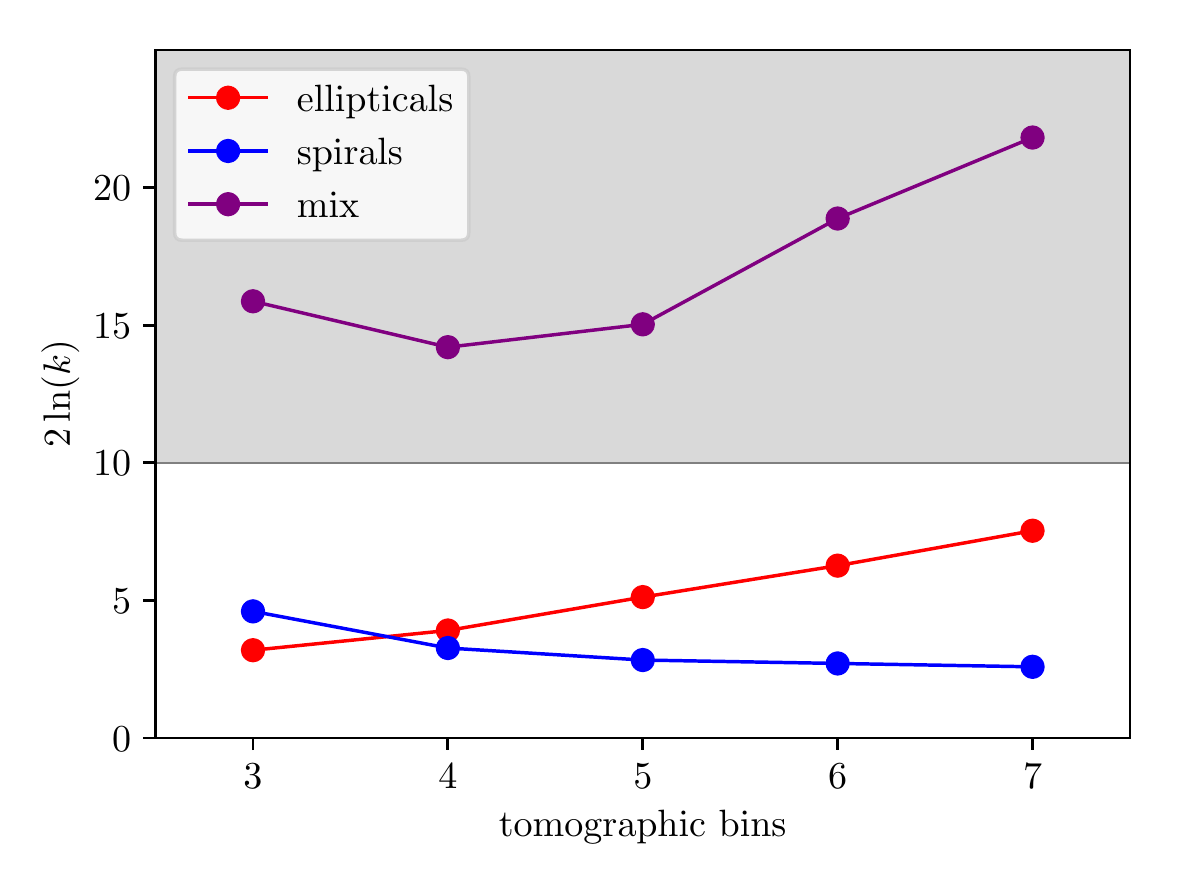}
\caption{Bayes factor $2\, \ln(k)$, obtained by integrating eq.~\ref{eq:BE}, for different biased weak lensing likelihoods $\likeli_f$, obtained from alignment models for $(i)$ only ellipticals, $(ii)$ only spirals, and $(iii)$ a mix of both. According to $2\, \ln(k)\geq10$ is interpreted as `very strong' evidence against the null hypothesis, in this case a $\Lambda$CDM cosmology.}
\label{fig:BE}
\end{figure}

This can be evaluated numerically for different priors and tomographic bin numbers, as shown in Fig.~\reff{fig:BE}. Our analysis shows that increasing the number has adverse effects on the individual alignment models, i.e. the tension between prior and shifted likelihood goes down for the quadratic alignment model, whereas in increases for the linear alignment model. Again, the results of a mix with spiral fraction $0.7$ does not seem to be a linear combination of both and their combined effects are consistently above $2\,\ln(k)=10$ on the Jeffrey's scale \citep[for a discussion about its applicability in cosmology, see][]{nesseris_is_2013}, which according to \cite{doi:10.1080/01621459.1995.10476572} would be interpreted as very strong incompatibility between the prior and the weak lensing measurement in the context of a $w$CDM-model due to the induced shift in the parameter set caused by the intrinsic alignment contamination. This effectively shows that ignoring the systematical effects of intrinsic alignments would lead to false conclusions about the tension of a $w$CDM cosmology with previous observations. But in this context a large bias in the dark energy equation of state $w$ would not seem to be possible, and a confusion between a dynamical dark energy component and a cosmological constant $\Lambda$ implausible.

% ---  --- %
\subsection{Kullback-Leibler-divergence introduced by systematics}
An alternative way of quantifying the mismatch between the original and shifted likelihood is the computation of the Kullback-Leibler divergence as a measure of the difference between two distributions.

In one of our previous investigations on the topic, we have defined the figure of bias 
\begin{equation}
Q^2 = \sum_{\mu\nu}F_{\mu\nu}\delta_\mu\delta_\nu, 
\end{equation}
which corresponds to half of the Kullback-Leibler-divergence under the assumption of two identical Gaussian likelihoods separated by a small bias shift $\delta_\mu$: We start at the general definition of the Kullback-Leibler divergence $D_\mathrm{KL}$,
\begin{equation}
D_\mathrm{KL} = \int \dd^N\theta\, \likeli_t(\vec{\theta}) \ln\left(\frac{\likeli_t(\vec{\theta})}{\likeli_f(\vec{\theta})}\right),
\label{eq:KLdef}
\end{equation}
which defines the logarithmic likelihood ratio between the true and the wrong model, integrated over the parameter space with the true likelihood as a weighting function.
Since the parameter space for both likelihoods is the same, we can re-write eq.~\ref{eq:KLdef} as 
\begin{equation}
%D_\mathrm{KL} = \int_\Theta \dd \likeli_t\, \ln\left(\frac{\dd \likeli_t}{\dd \likeli_f}\right) = \int \dd \likeli_f \frac{\dd \likeli_t}{\dd \likeli_f} \ln\left(\frac{\dd \likeli_t}{\dd \likeli_f}\right),
D_\mathrm{KL} = \int \dd^N\theta\, \likeli_t \ln \likeli_t - \int \dd^N\theta\, \likeli_t \ln \likeli_f,
\label{eq:KLentro}
\end{equation}
which is by definition the difference between the entropy of $\likeli_t(\vec{\theta})$ and the cross-entropy of $\likeli_f(\vec{\theta})$ and $\likeli_t(\vec{\theta})$. Therefore, $D_\mathrm{KL}$ is a measure for the relative entropy between the two models.

Assuming $N$-dimensional Gaussian likelihoods described by covariance matrices $C_a$, $a \in \{t,f\}$,
\begin{equation}
\likeli_a(\vec{\theta}) = 
\frac{1}{\sqrt{(2\pi)^N\det C_a}}
\exp\left(
-\frac{1}{2}
\left(\vec{\theta}-\vec{\theta}_a\right)_\mu 
\left(C_a^{-1}\right)_{\mu\nu} 
\left(\vec{\theta}-\vec{\theta}_a\right)_\nu
\right),
\end{equation} 
for the two likelihoods, it is clear that the distributions are separated by $\vec\delta = \vec\theta_{f}-\vec\theta_{t}$. If both $\likeli_t$ and $\likeli_f$ are Gaussian, the Kullback-Leibler divergence becomes
\begin{equation}
D_\mathrm{KL} = 
\frac{1}{2} \left( 
\trace(\,C_{f}^{-1} C_{t} ) + \sum_{\mu\nu}\left(C^{-1}_f\right)_{\mu\nu}\delta_\mu\delta_\nu - N + \ln\left(\frac{\det C_{f}}{\det C_{t}}\right)  
\right).
\end{equation}
With only a small shift $\delta_\mu$ in each parameter, it is feasible to set $C_t \simeq C_f$ and obtain the covariance from the Fisher matrix of the true model, $C_t = F^{-1}$. Consequently, eq.~\ref{eq:KLdef} reduces to 
\begin{equation}
D_\mathrm{KL} = \frac{1}{2} \left( N +  \sum_{\mu\nu}F_{\mu\nu}\delta_\mu\delta_\nu - N + \ln\ 1 \right) = \frac{Q^2}{2}.
\end{equation}
Therefore, the Kullback-Leibler divergence reduces to half of the figure of bias $Q^2$, as introduced in \cite{schaefer_angular_2015}, in this limit. Note that $\exp(-D_\mathrm{KL})$ is the ratio between the likelihood evaluated at a distance $\delta_\mu$ away from the best fit point without systematics relative to the original value. Intuitively, this means $Q^2$ or $D_\mathrm{KL}$ are quantifying the scale of the bias in relation to the statistical errors.

The preferred unit to measure the Kullback-Leibler divergence is usually nat, quantifying the information or entropy using the natural logarithm. In the langauge of statistical physics, this is equivalent to $S = k_{B} \ln \Omega $ with $k_{B} = 1$.

\begin{figure}
\includegraphics[width = \columnwidth]{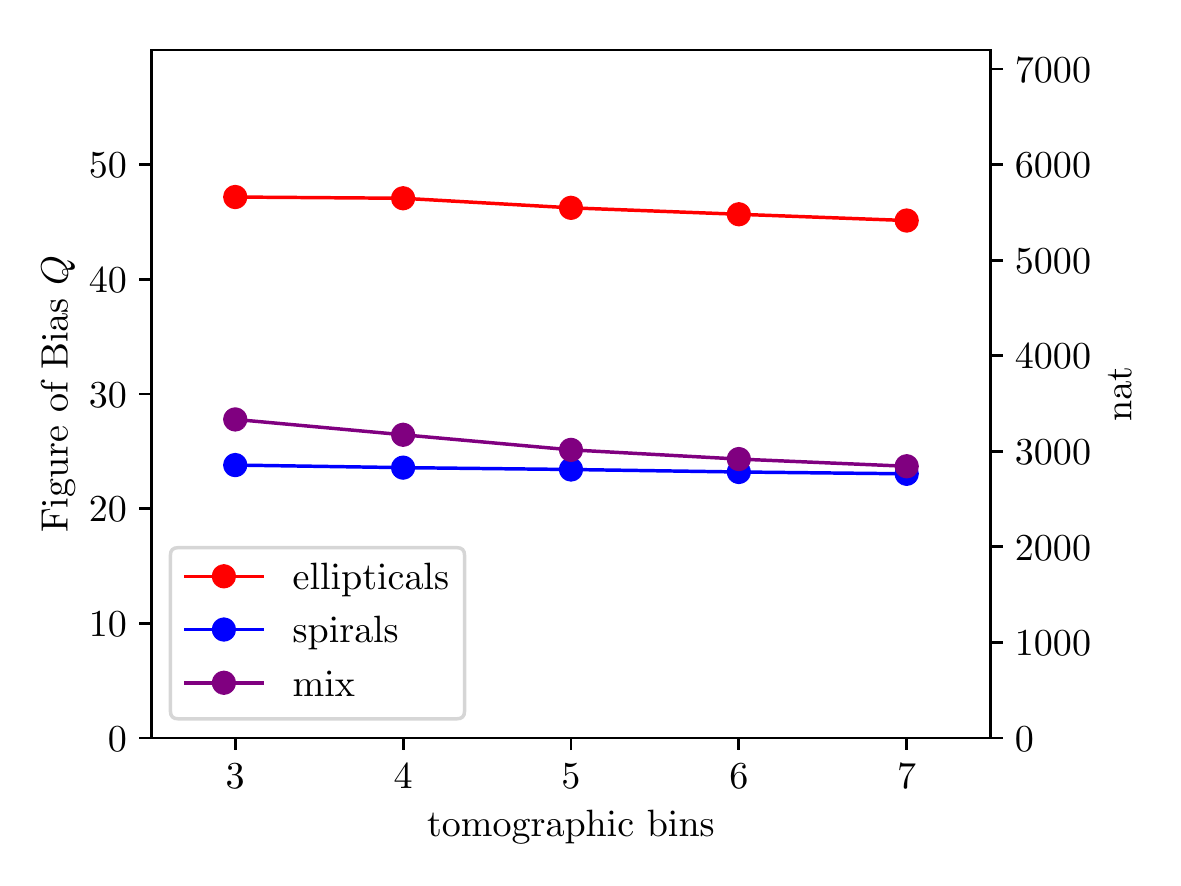}
\caption{Figure of bias $Q^2 = \sum_{\mu\nu}F_{\mu\nu}\delta_\mu\delta_\nu$ and Kullback-Leibler divergence $D_\mathrm{KL} = Q^2/2$ in nats, quantifying the bias in cosmological parameter estimation introduced by ignoring the intrinsic alignments signal of $(i)$ only ellipticals, $(ii)$ only spirals, and $(iii)$ a mix of both, in relation to their statistical errors.}
\label{fig:KL}
\end{figure}

For the spiral galaxy alignment bias only, we confirm the findings of \cite{schaefer_angular_2015}, namely $Q\simeq20$ and weak to no dependence of the number of tomographic bins, see~\reff{fig:KL}. For the bias caused by elliptical galaxies, the absolute number is a bit higher, albeit with the same overall beaviour. A mix of both types tends to be close to the spirals, possibly because of their dominance in the galaxy mix.

In comparison with the Bayes-factor (i.e. Fig.~\ref{fig:BE}, computed previously), the hierarchy of the different errors in the various models seems to be at odds at first glance. There, the mix of both models gives the most tension within a $w$CDM model class, whereas here, the ellipticals produce the largest figure of bias. This is alleviated by the fact that in the Bayes-factor a particular prior is included whereas this information is missing in the computation of the figure of bias $Q$ or the Kullback-Leibler divergence: The Bayes-factor quantifies the overlap of the biased likelihood with a prior distribution, which in turn specifies less favoured bias directions, i.e. those perpendicular to the degeneracy of the prior, cf. Fig.~\ref{fig:Fisher_BE}.

% ---  --- %
\section{Conclusions}\label{sect_conclusion}
The subject of this investigation were intrinsic ellipticity correlations of galaxies and their impact on the parameter estimation process from future tomographic weak lensing data. Intrinsic alignments were computed over the entire relevant multipole range using physically motivated models for spiral and elliptical galaxies, and used to predict tomographic angular spectra of the intrinsic ellipticity field. Specifically, we considered the case of parameter estimation based on the tomographic weak shear $E$-mode spectrum as measured by Euclid.

\begin{enumerate}
\item{Ellipticity correlations for spiral galaxies and elliptical galaxies result in our model from the tidal torquing and from the tidal shearing mechanism, respectively, and we used the same formalism for computing both alignments for a consistent modelling, and reduced ellipticity correlations to correlations in the tidal gravitational field. The model parameters in both cases have a clear physical interpretation: In the case of spiral galaxies, we assume that the observed ellipticity follows from the orientation of the angular momentum direction relative to the line of sight, and we link the angular momentum direction to the (squared) tidal shear field through a misalignment parameter $A$, whose value is derived from numerical simulations. Conversely, in elliptical galaxies the observed shape is a physical deformation of a virialised stellar system, and should be proportional ot the magnitude and orientation of the tidal shear, and we estimate this constant of proportionality $D$ from CFHTLenS's observation of intrinsic shape correlations in elliptical galaxies.}
\item{Ellipticity correlation functions for elliptical galaxies are obtained through Limber-projection of the tidal shear correlations restricted to components perpendicular to the line of sight: These result in angular ellipticity correlation functions $C(\theta)$ with correlation lengths between $10$ and $100$ arc-minutes for a galaxy survey like Euclid's, which is notably more long-ranged than correlations in the shapes of spiral galaxies which only exist on very small angular scales between $1$ and $10$ arc-minutes for an identical geometry. Consequently, angular ellipticity spectra for the shapes of elliptical galaxies are largest on intermediate multipoles below $\ell=1000$, while spiral galaxies show the largest amplitudes only on small multipoles above $\ell=1000$. In all cases, the amplitude of intrinsic shape correlations is smaller than that of the weak lensing prediction by about an order of magnitude, depending on the source redshifts. For small redshifts and $\ell\simeq 100$, the IA signal is of the same order as the weak lensing, whereas for high redshifts, it can be almost two orders of magnitude smaller.}
\item{Apart from II-correlations of both spiral and elliptical galaxies, there are nonzero cross-correlations between the intrinsic alignment of elliptical galaxies and weak gravitational lensing. Under the assumption of Gaussian statistics of tidal shear field and a linear response of elliptical galaxies to external tidal gravitational fields, one would not expect this effect for spiral galaxies, and likewise, there should be no intrinsic cross-alignment between the shapes of spiral and elliptical galaxies. The GI-signal is negative and dominates much of the high multipole range, while being rather small on intermediate multipoles. For combining the two alignment models we assume a morphological mix of galaxies with a fraction of $q=0.7$ of spiral galaxies and a fraction of $1-q=0.3$ of elliptical galaxies. The ellipticals show in addition a GI-cross alignment, whereas gravitational lensing is universal and does not differentiate between galaxy morphologies.}
\item{There is a great deal of uncertainty in the intrinsic alignment model parameters, and we were very conservative in assuming numerical values. For spiral galaxies, we adopted values for the average misalignment between the angular momentum direction and the tidal shear orientation of $A\simeq 0.25$. The effect of a finite thickness of galactic discs, which have been measured in spiral galaxies in the local Universe, is likewise absorbed into this constant. Conversely, we have estimated the constant of proportionality between ellipticity and tidal shear in elliptical galaxies from the weak alignment signal in CFHTLenS, and found values compatible with other studies. Specifically, we set up an estimate of the statistical significance of the intrinsic alignment signal in the presence of cosmic variance generated by weak lensing for a survey with CFHTLenS's characteristics, and obtained a value of $D\simeq 10^{-4}c^2$. It is remarkable that this constant of proportionality should, for a virialised elliptical galaxy, not scale with mass $M$, and the response of an elliptical galaxy to a tidal shear field is fast in comparison to the time scale of structure formation, so that there is no dependence on redshift $z$ either. We restrained from marginalising over different values of $D$ and $A$ in this work, as the impact of a fixed contamination of a Euclid-like survey with IAs was to be examined. Since they are physical parameters that are intrinsic to the galaxy morphologies and their alignment behaviour, we don't expect these parameters to vary wildly between different surveys. Only gross misclassifications of galaxy morphologies in these surveys would change the values of $A$ and $D$ drastically and make them and hence the used models more heuristic than physical in their nature.}
\item{While our statistical analysis focused on inference from the $E$-mode spectra of the observed ellipticity fields, the alignment models predict nonzero $B$-modes the II-correlation of all galaxies and for the GI-correlation for elliptical galaxies. This latter contribution is in fact dominating over almost the entire multipole range over $B$-mode generating effect in weak lensing in second order. We expect considerable potential for controlling intrinsic alignment contamination in weak lensing data by making use of the intrinsic $B$-modes and are in the process of investigating this issue. Ideally, one could estimate the spectra $C^{\epsilon,\mathrm{II}}_{B,ii}(\ell)$ and $C^{\epsilon,\mathrm{GI}}_{B,ij}(\ell)$ of the ellipticity field, which should be possible at high significance with e.g. Euclid-data \citep{schaefer_angular_2015} and use the relations between $E$- and $B$-mode amplitudes predicted by the models to control the $E$-mode contamination. \cite{blazek_beyond_2017} find as well that individual intrinsic alignment model parameters can be individually constrained and that all contributions to the ellipticity correlation are statistically significant for a survey like LSST\footnote{\url{https://www.lsst.org}}.}
\item{Intrinsic alignments of elliptical galaxies are, identically to gravitational lensing, sensitive to  fluctuations in the gravitational potential that scale in leading order with the product $\Omega_\mathrm{m}\times\sigma_8$. The scale-dependence of both effects is determined by the shape of the CDM-spectrum as fixed at lowest order by $\Omega_\mathrm{m}$ and $h$. These relationships are markedly different in spiral galaxies, where alignment is an orientation effect that does not depend on the strength of the potential fluctuations themselves, and on the shape of the spectrum through spectral moments that differ from those relevant for elliptical galaxies. All effects share a similar dependence on the dark energy properties as it enters through the Hubble-function $H(a)$ in converting redshifts to comoving distances at the stage of carrying out line of sight-projections. The biases we compute, in particular for the $\Omega_\mathrm{m}$-$\,\sigma_8$-plane, could not be of importance in explaining the tension between weak lensing surveys and CMB measurements of those parameters \citep[e.g. in][]{2016arXiv161004606J} as they would shift the likelihood into the wrong direction, further away from the CMB-constraints - this would primarily concern the linear alignment model of elliptical galaxies, because CFHTLenS would be insensitive to the alignment of spiral galaxies on high multipoles $\ell$ for our choice of model parameters. The situation would be different, however, if there was a strong IA-contribution from spiral galaxies on small angular scales, which could shift the likelihoods in a direction that would make them compatible with those from the CMB.}
\item{With the intrinsic alignment contamination of elliptical galaxies on intermediate multipoles, the cross-correlation between alignments and gravitational lensing on the full multipole range and spiral galaxies on high multipoles, we simulated the parameter estimation process and propagated the effect of the alignment contribution in the spectra to the systematic error in the estimates of cosmological parameters, if all alignments are neither removed nor modelled, and we compared this systematic error with the expected statistical precision which is derived with the Fisher-matrix formalism. Together with corrections to the matter spectrum due to baryonic physics \citep{white_baryons_2004, semboloni_quantifying_2011, merkel_parameter_2017}, intrinsic alignments are the largest contribution of systematic uncertainty to weak lensing data and dominate over second-order effects in lensing, which remains to be true for $B$-mode spectra.}
\item{We explicitly calculated the induced biases on five cosmological parameters, namely $\left\{\Omega_\mathrm{m},\sigma_8,h,n_s,w\right\}$ for a Euclid-like survey. Furthermore, we investigated the behaviour of these biases with increasing bin number. As expected, the largest error is made on those parameters to which weak lensing is most sensitive, $\Omega_\mathrm{m}$ and $\sigma_8$. For these, we see systematical errors on the order of two times the statistical uncertainty for at least one of our models. In addition, the error on the spectral index $n_s$ is heavily affected, in particular by a mix of the two models, possibly because the shape of the spectrum is not much altered by additional power over a limit range in $\ell$ from only one model, but can be altered more heavily by a combination of two contributions at different multipoles. The Hubble parameter $h$ and the equation of state $w$ seem least affected by the omission of IA in the parameter inference: This is due to the fact that in comparison to other parameters the lensing signal is least sensitive to them. It is difficult to compare forecasts of parameter biases because they not only depend on the intrinsic alignment model but as well on the methodology, the choice of the experimental setup that is simulated, the considered parameter space and possible inclusion of other observables and priors. \citep{kirk_cosmological_2012} find highly significant biases from a linear alignment model on the dark energy equation of state parameters and shows that they can be controlled by introducing a flexible IA-modelling at the expense of statistical accuracy. Our analysis for the linear alignment model is in agreement with \cite{kirk_impact_2010} concerning the values of $\sigma_8$ which are biased towards lower values and could only expain an increase in $\Omega_\mathrm{m}$ with the inclusion of spiral galaxies. In a similar study for LSST, \citet{blazek_beyond_2017} compute the impact of a combined alignment model on cosmological parameters and points out the importance of a correct intrinsic alignment model, in particular if one restricts the model to a purely linear one for elliptical galaxies. Comparing surveys such as Euclid and to ones such as LSST, it is clear that space-based surveys, being broader in area yet shallower in redshift, will be affected more by intrinsic alignments than their ground-based counterparts: Lensing efficiency (peaking at medium redshift) changes how the locally constant signal amplitude of the IAs relates to lensing and as the statistical error bars shrink by covering more galaxies and scales, the (relatively constant) biases grow in relation. Therefore, space-based surveys are more prone to produce biases in the inferred cosmological parameter set due to IAs.}
\item{Apart from providing systematical errors in the full set of parameter for a dark energy cosmology, we quantified the contaminating effect of intrinsic alignments from the point of view of the Bayes-evidence and the Kullback-Leibler divergence. If there are systematic errors in the estimated parameter set, it will be the case that the evidence for a certain model, for instance $\Lambda$CDM, decreases because the contaminated weak lensing likelihood is incompatible to some degree with the prior knowledge on the parameter set. We have computed the loss in Bayesian evidence for a $w$CDM-model due to intrinsic alignments in weak lensing data if the prior is given through high-resolution CMB-observations. We found a significant loss of evidence in favour of a specific $w$CDM-parameter set in terms of the Jeffreys-scale; in fact, neglecting the IA contributions would make it seem like a $w$CDM model inferred from weak lensing would be completely incompatible with a Planck-like CMB prior. In a similar vein, we quantified the incompatibility between the contaminated lensing likelihoods with the Kullback-Leibler divergence as a measure of the difference between the true and the wrongly inferred likelihood. In comparison to the previously defined figure of bias $Q^2$ we were able to show that this particular number corresponds to the Kullback-Leibler divergence in the case of constant covariance matrices, and confirmed previously obtained numbers \citep{schaefer_angular_2015}. It should be emphasised, however, that the reverse, i.e. mistaking a dark energy model for a model with a cosmological constant, is very unlikely, as the induced biases are smaller than the statistical error in the dark energy equation of state.}
\end{enumerate}

A logical next step to our investigation would be a separation of the different alignment types on the basis of their different physical and statistical characteristics and the application of mitigation techniques that differentiate between galaxy morphologies. This, in particular, requires corrections to the tidal shear fields due to nonlinear evolution of the cosmic large-scale structure and to go beyond the linear and Gaussian regime of structure formation, which allows then cross-correlations between the shapes of elliptical and spiral galaxies, and an non-zero associated GI-effect for spiral galaxies. Secondly, as our statistical investigation entirely depended on the assumption of Gaussian statistics, it would be very important to extend the investigation to Monte-Carlo Markov-chain evaluated, biased likelihoods.

Furthermore, a compelling method of simplifying the problems that arise from different alignment processes would be to separate the galaxy types observationally and test the models against the individual results. There are a few challenges connected with this: Firstly, one would lose significant statistical power by only taking a subset of the available data. Then, the separation into two different types of galaxies cannot be done with full confidence by morphology, (intrinsic) colour or environment alone. The safest way would be to use spectroscopic data as well, which is unavailable for the majority of the galaxies e.g. in the Euclid survey. However, a combination of all of those differentiators, possibly restricted to low-to-mid redshifts, could lead to a viable result until more reliable ways of distinguishing galaxy types are feasible.

% ---  --- %
\section*{Acknowledgements}
TMT acknowledges financial support from a doctoral fellowship by the Astronomisches Rechen-Institut, Heidelberg. We thank R. Reischke for providing a Fisher-matrix for CMB-spectra and for galaxy clustering. We would like to thank Ph. Umst{\"a}tter for sharing his results on the linear response of haloes to tidal gravitational fields, and Ph.M. Merkel for valuable comments. BMS appreciates conversations with D. Piras and B. Joachimi on the alignment processes in elliptical galaxies.

% ---  --- %
\bibliographystyle{mnras}
\bibliography{references}

% ---  --- %
\appendix

\bsp
\label{lastpage}
\end{document}